\documentclass[journal]{IEEEtran}

\usepackage{graphicx}
\usepackage{amsmath,epsfig,amssymb, amstext, verbatim,amsopn,cite,subfigure,multirow,multicol,lipsum}
\usepackage{balance}
\usepackage{url}
\usepackage{amsfonts}
\usepackage{epsfig}
\usepackage{slashbox}
\usepackage{setspace}
\usepackage{stmaryrd}
\usepackage{psfrag}	
\usepackage{multirow}
\usepackage{float}
\usepackage{slashbox}
\usepackage[process=auto]{pstool}
\usepackage{etoolbox}
\allowdisplaybreaks

\newtoggle{paper}
\togglefalse{paper}

\iftoggle{paper}{
  
}{
  
}

\newcommand{\Equal}{\hspace{-0.3mm}=\hspace{-0.3mm}}
\newcommand{\Add}{\hspace{-0.3mm}+\hspace{-0.3mm}}
\newcommand{\Minus}{\hspace{-0.3mm}-\hspace{-0.3mm}}
\newcommand{\Less}{\hspace{-0.3mm}<\hspace{-0.3mm}}
\newcommand{\Great}{\hspace{-0.3mm}>\hspace{-0.3mm}}
\newcommand{\Lequal}{\hspace{-0.3mm}\leq\hspace{-0.3mm}}
\newcommand{\Gequal}{\hspace{-0.3mm}\geq\hspace{-0.3mm}}
\newcommand{\Nequal}{\hspace{-0.3mm}\neq\hspace{-0.3mm}}

\newtheorem{lem}{Lemma}
\newtheorem{remk}{Remark}

\newtheorem{Prop}{Proposition}

\EndPreamble
\begin{document}
\pagenumbering{gobble} 

\title{A Delay-Constrained Protocol with Adaptive Mode Selection for Bidirectional  Relay Networks \vspace{-0.4cm}}
\author{Vahid Jamali$^\dag$, Nikola Zlatanov$^\ddag$, and Robert Schober$^\dag$ \\
\IEEEauthorblockA{$^\dag$ Friedrich-Alexander-University Erlangen-N\"{u}rnberg (FAU), Germany \\
 $^\ddag$ University of British Columbia (UBC), Vancouver, Canada}
\vspace{-1.1cm}
\thanks{This technical report is an extended version of a paper submitted to IEEE Globecom 2014.}}

\maketitle

\begin{abstract}
In this paper, we consider a bidirectional relay network with half-duplex nodes and block fading where the nodes transmit with a fixed transmission rate. Thereby, user 1 and user 2 exchange information
only  via  a  relay  node,  i.e.,  a  direct  link  between  both  users  is
not  present.  Recently in \cite{ICCIEEE}, it was shown that a considerable gain in terms of sum throughput can be obtained in bidirectional relaying by optimally selecting the transmission modes or, equivalently, the states of the nodes, i.e., the transmit, the receive, and the silent states, in each time slot based on the qualities of the involved links. To enable adaptive transmission mode selection, the relay has to be equipped with two buffers for storage of the data received from the two users. However, the  protocol proposed in \cite{ICCIEEE} was delay-unconstrained and provides an upper bound for the performance of practical delay-constrained protocols. In this paper, we propose a heuristic but efficient delay-constrained protocol which can approach the performance upper bound reported in \cite{ICCIEEE}, even in cases where only a small average delay is permitted.
In particular, the proposed protocol does not only take into account the instantaneous qualities of the involved links for adaptive mode selection but also the states of the queues at the buffers. The average throughput and the average delay of the proposed delay-constrained protocol are evaluated by analyzing the Markov chain of the states of the queues. 
\end{abstract}

\section{Introduction} \label{Sec I (Intro)}

In the bidirectional relay network, two users  exchange information via a relay node. For this simple and fundamental network architecture, several protocols have been
proposed for practical half-duplex nodes,  i.e.,  nodes that  cannot  transmit  and  receive  at  the
same  time  and  in  the  same  frequency  band. The traditional two-way relaying protocol,   the time division broadcast (TDBC)  protocol \cite{TDBC}, and the multiple access
broadcast (MABC) protocol \cite{MABC} are the most widely used protocols for the bidirectional relay channel. For a comprehensive overview of protocols proposed for the bidirectional relay channel, we refer to \cite{Tarokh,BocheIT,TDBC,MABC}, and references therein. Notice that  the protocols in \cite{Tarokh,BocheIT,TDBC,MABC}  were  derived
for adaptive rate transmission which requires the availability
of  channel  state  information (CSI) at all transmitting nodes  and
the  capability  of  using  appropriate  coding  and  modulation
schemes such that the transmitters can perfectly adapt their transmission
rates to the channel capacity. For the case when CSI is not
available at all transmitting nodes and/or only one coding and modulation scheme can be
used,  protocols  designed for  adaptive  rate  transmission  are  not
applicable. Instead, the transmitters have to transmit with a fixed
rate regardless of the CSI of the
involved links. For fixed rate transmission, not the achievable
rates but other performance metrics such as throughput
and outage probability are relevant \cite{ICCIEEE,Outage2}.

Most of the previous bidirectional relaying protocols assume a prefixed schedule for the nodes to transmit, receive, and be silent, i.e., a fixed schedule for using the possible transmission modes listed in Table I. In \cite{ICCIEEE}, a new protocol for fixed rate transmission is proposed which, based on the qualities of the involved links, selects the optimal transmission mode in each time slot such that the sum throughput is maximized. However, the protocol proposed in \cite{ICCIEEE} does not impose any constraint on the average delays of the information flows and may lead to unlimited average delays. Nevertheless, for most practical applications, it is required that the end-to-end delay does not exceed a certain tolerable limit. Hence, in this paper, we propose a delay-constrained protocol which can guarantee a certain target average delay for each information flow. In particular, the proposed delay-constrained protocol does not only take into account the qualities of the links for adaptive mode selection, but also the states of the queues at the buffers. Thereby, the proposed protocol avoids excessive delays by effectively forcing the relay to transmit if the amount of information in the queues exceeds a certain threshold. 

\begin{table}
\label{Modes}
\caption{Transmission Modes for the Considered Bidirectional Relay Network (T: Transmit, R: Receive, S: Silent).}  \vspace{-0.4cm}
\begin{center}
\scalebox{0.8}{
\begin{tabular}{|| c | c | c | c |c | c | c | c ||}
  \hline                  
\textbf{Transmission Mode} & $\mathcal{M}_1$ & $\mathcal{M}_2$ & $\mathcal{M}_3$ & $\mathcal{M}_4$ & $\mathcal{M}_5$ & $\mathcal{M}_6$ & $\mathcal{M}_7$ \\ \hline
\textbf{User 1} & T & S & T & R & S & R & S \\ \hline
\textbf{User 2} & S & T & T & S & R & R & S \\ \hline
\textbf{Relay} & R & R & R & T & T & T & S \\ \hline
  
\end{tabular}
}
\end{center}
\vspace{-0.6cm}
\end{table}

The proposed protocol can operate in two modes: \textit{i)} a delay-efficient mode for stringent average delay requirements, and \textit{ii)} a throughput-efficient mode for less stringent average delay requirements. 
For performance analysis of the proposed protocol, we present a general framework for obtaining the average throughput and the average delay of each information flow based on a Markov chain analysis of the states of the queues at the buffers.  The performance analysis reveals that the signal-to-noise ratio (SNR) gap between the outage probability of the proposed protocol in the delay-efficient mode and that of the delay-unconstrained protocol in \cite{ICCIEEE} is at most $3$ dB in the high SNR regime. Furthermore, the SNR gap  between the proposed protocol in the throughput-efficient mode and the delay-unconstrained protocol vanishes at high SNRs.

\iftoggle{paper}{
  
}{
We note that buffer-aided relaying has been considered in the literature for different network architectures, e.g., the one-way relay network \cite{NikolaMixed,Poor,Ding}, the two-way relay network \cite{ITArxiv,EUSIPCOIEEE,PopovskiLetter}, the multihop relay network \cite{HanzoMultihop}, and the diamond relay network \cite{Aissa,Thompson}. In particular, for the one-way relaying,  a delay-constrained protocol for fixed rate transmission was proposed in \cite{NikolaMixed}.   
}

\section{System Model and Preliminaries}\label{SysMod}

In this section, we  introduce the system model and  present some preliminaries for development of the proposed protocol.

\subsection{System Model}

We consider a bidirectional relay network comprised of two users and a relay. There is no direct link between the users, and thus, user 1 and user 2 communicate with
each other only through the relay node, see Fig. 1. All
three nodes in the network are assumed to be half-duplex.  Moreover, the user-to-relay and relay-to-user channels are impaired
by additive white Gaussian noise (AWGN) and block fading, i.e., the channel 
coefficients are constant during one time slot and change independently from
one time slot to the next. Let $h_1(i)$ and $h_2(i)$ denote the channel fading coefficients between user 1 and the relay and between user 2 and the relay in the $i$-th time slot, respectively. Fading gains $|h_1(i)|^2$ and $|h_2(i)|^2$ are assumed to be ergodic and stationary random processes with means $\Omega_1=E\{|h_1(i)|^2\}$ and  $\Omega_2=E\{|h_2(i)|^2\}$, respectively, where $E\{\cdot\}$ denotes expectation. 
Moreover, $\gamma_1(i)=\gamma|h_1(i)|^2$ and $\gamma_2(i)=\gamma|h_2(i)|^2$ denote the instantaneous SNRs of the links between user 1 and the relay and user 2 and the relay, respectively, where $\gamma=\frac{P}{\sigma_n^2}$ is the transmit SNR of the nodes, $P$ is the transmit power of the nodes, and $\sigma_n^2$ is the noise variance at the receivers.  We also assume that all nodes transmit one packet in each time slot with fixed rate $R_0$. 

\begin{figure}
\centering
 \pstool[width=0.9\linewidth]
{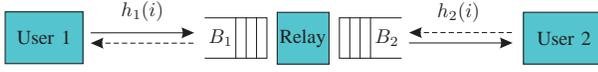}{
\psfrag{U1}[c][c][0.75]{$\text{User 1}$}
\psfrag{U2}[c][c][0.75]{$\text{User 2}$}
\psfrag{R}[c][c][0.75]{$\text{Relay}$}
\psfrag{h1}[c][c][0.75]{$h_1(i)$}
\psfrag{h2}[c][c][0.75]{$h_2(i)$}
\psfrag{B1}[c][c][0.75]{$B_1$}
\psfrag{B2}[c][c][0.75]{$B_2$}} \vspace{-0.3cm}
\caption{Bidirectional relay network consisting of two users and a buffer-aided relay.}
\label{FigSysMod}
\vspace{-0.6cm}
\end{figure}

\subsection{Instantaneous SNR and Queue Regions}

Let $B_1$ and $B_2$ denote two buffers at the relay which store the  information received from user 1 and user 2, respectively. Moreover, $\ell_j(i), \,\, j\in\{1,2\}, \,\ell_j=0,\dots,\ell_j^\mathrm{max}$, denotes the number of packets available in buffer $B_j$ at the end of the $i$-th time slot. In order to avoid information loss, transmission mode ${\cal M}_k$ is selected only if 1) the information can be decoded successfully at the receiver(s) based on the qualities of the respective links, and 2) there is enough space available in the respective buffer(s) to store data for modes $\mathcal{M}_1$, $\mathcal{M}_2$, and $\mathcal{M}_3$, or there is enough information available in the respective buffer(s) to transmit for modes $\mathcal{M}_4$, $\mathcal{M}_5$, and $\mathcal{M}_6$.  Otherwise, the silent mode $\mathcal{M}_7$ is selected. 

\begin{figure}
\centering
 \pstool[width=1\linewidth] 
{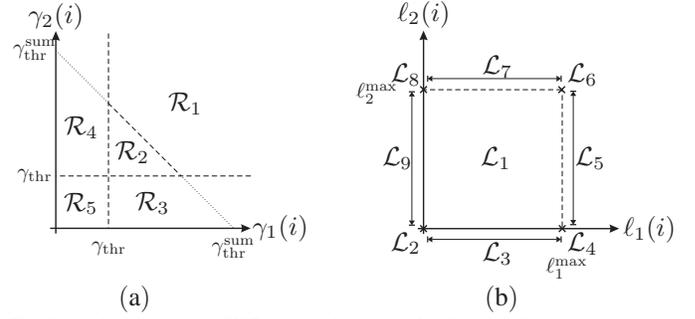}{
\psfrag{R1}[c][c][1]{$\mathcal{R}_1$}
\psfrag{R2}[c][c][1]{$\mathcal{R}_2$}
\psfrag{R3}[c][c][1]{$\mathcal{R}_3$}
\psfrag{R4}[c][c][1]{$\mathcal{R}_4$}
\psfrag{R5}[c][c][1]{$\mathcal{R}_5$}
\psfrag{G1}[c][c][1]{$\gamma_1(i)$}
\psfrag{G2}[c][c][1]{$\gamma_2(i)$}
\psfrag{Gthr}[c][c][0.8]{$\gamma_{\mathrm{thr}}$}
\psfrag{Gsum}[c][c][0.8]{$\gamma_{\mathrm{thr}}^{\mathrm{sum}}$}
\psfrag{Q1}[c][c][1]{$\mathcal{L}_1$}
\psfrag{Q2}[c][c][1]{$\mathcal{L}_2$}
\psfrag{Q3}[c][c][1]{$\mathcal{L}_3$}
\psfrag{Q4}[c][c][1]{$\mathcal{L}_4$}
\psfrag{Q5}[c][c][1]{$\mathcal{L}_5$}
\psfrag{Q6}[c][c][1]{$\mathcal{L}_6$}
\psfrag{Q7}[c][c][1]{$\mathcal{L}_7$}
\psfrag{Q8}[c][c][1]{$\mathcal{L}_8$}
\psfrag{Q9}[c][c][1]{$\mathcal{L}_9$}
\psfrag{l1}[c][c][1]{$\quad\ell_1(i)$}
\psfrag{l2}[c][c][1]{$\ell_2(i)$}
\psfrag{Qmax1}[c][c][0.8]{$\ell_1^{\max}$}
\psfrag{Qmax2}[c][c][0.8]{$\ell_2^{\max}$}
\psfrag{a}[r][c][1]{$(\text{a})$}
\psfrag{b}[r][c][1]{$(\text{b})$}} \vspace{-0.8cm}
\caption{a) Instantaneous SNR regions based on the decodability of information at the receivers in the different transmission modes, and b) instantaneous queue regions based on whether the queues of the buffers are empty, partially full, or completely full.}
\label{FigOutQReg}
\vspace{-0.5cm}
\end{figure}

 Fig. \ref{FigOutQReg} a) illustrates the five SNR regions, $\mathcal{R}_m,\,\,m=1,\dots,5$,  for the instantaneous link SNRs, $\boldsymbol{\gamma}(i)=[\gamma_1(i),\gamma_2(i)]$, that can be distinguished based on the decodability of information at the receivers. The boundaries of the SNR regions in Fig. \ref{FigOutQReg} a) are defined by $\gamma_{\mathrm{thr}}=2^{R_0}-1$ and $\gamma_{\mathrm{thr}}^{\mathrm{sum}} = 2^{2R_0}-1$. Moreover, for future reference, let $\mathcal{K}_{\mathcal{R}_m},\,\,m=1,\dots,5$ denote the set of the indexes of the candidate transmission modes in SNR region $\mathcal{R}_m$, i.e., $\mathcal{K}_{\mathcal{R}_1}=\{1,\dots,7\}$, $\mathcal{K}_{\mathcal{R}_2}=\{1,2,4,5,6,7\}$, $\mathcal{K}_{\mathcal{R}_3}=\{1,4,7\}$, $\mathcal{K}_{\mathcal{R}_4}=\{2,5,7\}$, $\mathcal{K}_{\mathcal{R}_5}=\{7\}$. On the other hand, in Fig. \ref{FigOutQReg} b), nine different queue regions, $\mathcal{L}_n,\,\,n=1,\dots,9$,  are defined for the instantaneous states of the queues, $\boldsymbol{\ell}(i)=[\ell_1(i),\ell_2(i)]$,  based on whether the buffers are empty, partially full, or completely full. Let $\mathcal{K}_{\mathcal{L}_n},\,\,n=1,\dots,5$ denote the set of candidate transmission modes based on the states of the queues $\mathcal{L}_n$, i.e., $\mathcal{K}_{\mathcal{L}_1}=\{1,\dots,7\}$, $\mathcal{K}_{\mathcal{L}_2}=\{1,2,3,7\}$, $\mathcal{K}_{\mathcal{L}_3}=\{1,2,3,5,7\}$, $\mathcal{K}_{\mathcal{L}_4}=\{2,5,7\}$, $\mathcal{K}_{\mathcal{L}_5}=\{2,4,5,6,7\}$, $\mathcal{K}_{\mathcal{L}_6}=\{4,5,6,7\}$, $\mathcal{K}_{\mathcal{L}_7}=\{1,4,5,6,7\}$, $\mathcal{K}_{\mathcal{L}_8}=\{1,4,7\}$, $\mathcal{K}_{\mathcal{L}_9}=\{1,2,3,4,7\}$.  To summarize, only the transmission modes from the following set can be selected in time slot $i$
\begin{IEEEeqnarray}{ll} \label{SetF} 
		\mathcal{F}=\Big\{k \in\mathcal{K}_{\mathcal{R}_m}\cap\mathcal{K}_{\mathcal{L}_n}\big| \boldsymbol{\gamma}(i)\in\mathcal{R}_m\,\,\wedge\,\, \boldsymbol{\ell}(i-1)\in\mathcal{L}_n \Big\}. \quad
\end{IEEEeqnarray}

\subsection{Mode Selection Variables}

For the development of the proposed adaptive mode selection protocols, it is convenient to introduce seven binary mode selection variables,  $q_k(i) \in\{0,1\}, \,\,k=1,\dots,7$, where  $q_k(i)=1$ if  mode $\mathcal{M}_k$ is selected and $q_k(i)=0$ if it is not selected in the $i$-th time slot. Furthermore, we assume that, in each time slot,  only one of the seven transmission modes can be selected, i.e., $\mathop \sum_{k = 1}^7 q_k(i)=1$ holds.  

\section{Proposed Delay-Constrained Protocol}

As shown in \cite{ICCIEEE}, for sum throughput maximization, the queues of the buffers have to be at the edge of non-absorption. However, in this case, the size of the queues may increase as the number of time slots, $N$, tends to infinity. The idea of the  delay-constrained protocol proposed in this paper is to operate the lengths of the queues in buffers $B_1$ and $B_2$ around certain values $\ell_1^{\mathrm{thr}}$ and $\ell_2^{\mathrm{thr}}$, respectively. Hence, by choosing sufficiently small $\ell_j^{\mathrm{thr}},\,\,j=1,2$,  the average delays can be limited to certain desired values. 

Before we formally present the protocol, we introduce the general delay-aware utility function $\Lambda_k(i)$ for  transmission mode $\mathcal{M}_{k}$ which is designed such that it favours the selection of modes $\mathcal{M}_1$ and  $\mathcal{M}_3$ if $\ell_1(i-1)< \ell_1^{\mathrm{thr}} $,  modes $\mathcal{M}_2$ and  $\mathcal{M}_3$ if $\ell_2(i-1) < \ell_2^{\mathrm{thr}} $, modes $\mathcal{M}_4$ and  $\mathcal{M}_6$ if $\ell_2(i-1)> \ell_2^{\mathrm{thr}} $, and modes $\mathcal{M}_5$ and  $\mathcal{M}_6$ if $\ell_1(i-1)> \ell_1^{\mathrm{thr}} $. Furthermore,  $\tau_k$  denotes a utility function representing the spectral efficiency of transmission mode $\mathcal{M}_{k}$.  Examples for possible choices of $\Lambda_k(i)$ and $\tau_k$ will be provided at the end of this subsection. The proposed protocol operates in one of two modes, \textit{i)} a delay-efficient mode for stringent average delay requirements, and \textit{ii)} a throughput-efficient mode for less stringent average delay requirements.  In particular, from  set  $\mathcal{F}$, 
the protocol for the delay-efficient mode first selects the best transmission modes based on the delay-aware utility function $\Lambda_k(i)$, i.e.,
\begin{IEEEeqnarray}{ll} \label{SetU_Delay} 
		\mathcal{U}^{\mathrm{d}}=\Big\{k \big|k=\arg \underset{k\in\mathcal{F}^\mathrm{d}}{\max}\,\, \Lambda_k(i) \Big\} 
\end{IEEEeqnarray}
where $\mathcal{F}^\mathrm{d}\Equal \mathcal{F}$. Then, from set $\mathcal{U}^{\mathrm{d}}$ of the remaining modes with identical values of $\Lambda_k(i)$, the protocol selects the modes with the highest spectrally efficiency based on $\tau_k$, i.e.,
\begin{IEEEeqnarray}{ll} \label{SetU_Thr}
		\mathcal{U}^{\mathrm{t}}=\Big\{k \big|k=\arg \underset{k\in\mathcal{F}^\mathrm{t}}{\max}\,\, \tau_k \Big\} 
\end{IEEEeqnarray}
where $\mathcal{F}^\mathrm{t}\Equal \mathcal{U}^\mathrm{d}$.  On the other hand, for the throughput-efficient mode, the protocol first selects the best transmission modes based on the utility functions $\tau_k$, and then from the remaining transmission modes with identical values of $\tau_k$, the protocol selects the best transmission modes based on $\Lambda_k(i)$. Furthermore, for both the delay-efficient and throughput-efficient modes, if there are multiple candidates with identical values of $\Lambda_k(i)$ and $\tau_k$, the protocol selects the final transmission mode from the candidate set with equal probabilities.  To model the final mode selection mathematically, we define $X_{\mathcal{U}}(i)\in \mathcal{U}$ as the outcome of rolling a die with $|\mathcal{U}|$ equiprobable faces in the $i$-th time slot where $|\cdot|$ denotes the cardinality of a set and $\mathcal{U}$ is the final set of final candidate modes. In the following, we propose the protocol in compact form.

\textit{Delay-Constrained Protocol:} For the considered half-duplex  bidirectional  relay  network, the following adaptive mode selection leads to delay-constrained transmission
\begin{IEEEeqnarray}{lll} \label{Prot1}  
		q_{k^*}(i) = \begin{cases}
		1,\quad &\mathrm{if} \,\, k^*=X_{\mathcal{U}}(i)\\
		0, & \mathrm{otherwise}
		\end{cases}
\end{IEEEeqnarray}
where, in (\ref{SetU_Delay}) and (\ref{SetU_Thr}), for the delay-efficient mode, we set $\mathcal{U}=\mathcal{U}^{\mathrm{t}}$, $\mathcal{F}^\mathrm{t}=\mathcal{U}^{\mathrm{d}}$, and $\mathcal{F}^\mathrm{d}=\mathcal{F}$, and for the throughput-efficient mode, we set $\mathcal{U}=\mathcal{U}^{\mathrm{d}}$, $\mathcal{F}^\mathrm{d}=\mathcal{U}^{\mathrm{t}}$, and $\mathcal{F}^\mathrm{t}=\mathcal{F}$.  Furthermore, $\ell_1^{\mathrm{thr}} $ and $\ell_2^{\mathrm{thr}}$ are constants and are chosen to guarantee certain target average delays $\bar{T}_1^{\mathrm{d}}$ and $\bar{T}_2^{\mathrm{d}}$, respectively.

\textit{Proposed Utility Functions:}
The delay-aware utility function $\Lambda_k(i)$ has to be designed such that the numbers of packets in  buffers $B_1$ and $B_2$ remain close to $\ell_1^{\mathrm{thr}} $ and $\ell_2^{\mathrm{thr}}$, respectively. For instance, if $\ell_1(i-1)<\ell_1^{\mathrm{thr}} $ and we have the choice to select between transmission modes $\mathcal{M}_1$ and $\mathcal{M}_5$, the proposed protocol should select mode $\mathcal{M}_1$ to fill the buffer. Moreover, the utility functions $\Lambda_k(i)$ have to take into account for the states of the queues in both buffers. For example, if $\ell_1(i-1)=0$ and $\ell_2(i-1)=3$ where $\ell_1^{\mathrm{thr}} =\ell_2^{\mathrm{thr}} = 5$, and we have the choice to select either mode $\mathcal{M}_1$ or $\mathcal{M}_2$, the adopted utility function must give priority to mode $\mathcal{M}_1$. Taking into account these considerations for all subsets of the seven possible transmission modes, there is a large number of possible utility functions  $\Lambda_k(i)$. In the following, we propose one set of utility functions which can effectively limit the average delays. The proposed delay-aware utility functions are  
\begin{IEEEeqnarray}{lll} \label{SelecMet_Prot1}  
		\Lambda_1(i) = \ell_1^{\mathrm{thr}} - \ell_1(i-1)  \IEEEyesnumber\IEEEyessubnumber \\
		\Lambda_2(i) = \ell_2^{\mathrm{thr}} - \ell_2(i-1)  \IEEEyessubnumber \\
		\Lambda_3(i) = \min\{\Lambda_1(i),\Lambda_2(i)\}  \IEEEyessubnumber \\
		\Lambda_4(i) = \left[\ell_2(i-1) - \ell_2^{\mathrm{thr}} \right]^+ \IEEEyessubnumber \\
		\Lambda_5(i) = \left[\ell_1(i-1) - \ell_1^{\mathrm{thr}} \right]^+  \IEEEyessubnumber \\
		\Lambda_6(i) = \max\{\Lambda_4(i),\Lambda_5(i)\}   \IEEEyessubnumber \\
		\Lambda_7(i) = 0  \IEEEyessubnumber 
\end{IEEEeqnarray}
where $[x]^+=\max\{x,0\}$. The motivation for the above utility functions is as follows. First,  we assign the value of zero for the silent mode as a reference. If the numbers of packets in the queues are below their respective thresholds, $\Lambda_1(i)$, $\Lambda_2(i)$, and $\Lambda_3(i)$ are positive which favours the selection of the user-to-relay transmission modes. On the other hand, if the numbers of packets in the queues exceed the respective thresholds, $\Lambda_4(i)$, $\Lambda_5(i)$, and $\Lambda_6(i)$ are positive which favours the selection of the relay-to-user transmission modes. The comparison between the point-to-point transmission modes is based on how much the numbers of packets in the queues deviate from their respective thresholds. Moreover, selecting the relay-to-user transmission modes leads to a decrease of the average  delay while selecting the user-to-relay transmission modes might lead to an increase of the average delay. Thus, we use  $[\cdot]^+$ for modes $\mathcal{M}_4$ and $\mathcal{M}_5$ and not for modes $\mathcal{M}_1$ and $\mathcal{M}_2$, and we use $``\max"$ for mode $\mathcal{M}_6$ and $``\min"$ for mode $\mathcal{M}_3$. 

Furthermore, since we assume that the nodes transmit with fixed rate $R_0$, the spectral efficiency of a transmission mode is defined as the number of packets that is transmitted during one time slot. Hence, we choose the utility functions $\tau_k$ as $\tau_3=\tau_6=2$, $\tau_1=\tau_2=\tau_4=\tau_5=1$, and $\tau_7=0$. 

\begin{remk}
We  assume  that  the  relay  is  responsible  for
performing the mode selection. In particular, in the beginning of each time slot,
the users send pilots to the relay. The relay has to determine
the instantaneous SNR region. Thereby, using also the states of the queues, the relay is able to select the
 transmission  mode  according  to the proposed protocol.  Then,
the  relay  broadcasts  the  optimal  transmission  mode  to  the
users  using  three  bits  of  feedback  and  transmission  begins.
\end{remk}

\section{Throughput-Delay Analysis}

\iftoggle{paper}{

In this section, we present a general framework for the throughput-delay analysis of any delay-constrained protocol with adaptive mode selection for the  bidirectional relaying network. Moreover, as an example, we use the framework to derive some performance results for the  protocol proposed in Section III. Due to space constraints, we only provide the results for the delay-efficient mode. The analysis of the  protocol for the  throughput-efficient mode is provided in a technical report \cite{GlobeCom2014Arxiv} which is an extended version of this paper.
  
}{
  
  In this section, we present a general framework for the throughput-delay analysis of any delay-constrained protocol with adaptive mode selection for the  bidirectional relaying network considered in this paper. Moreover, as an example, we use the framework to derive some performance results for the proposed protocol.
  
}

\subsection{General Analysis of Adaptive Mode Selection Protocols}

Let $\bar{R}_{jj'}$ denote the average number of information bits/symbol received at node $j'$ from node $j$. Moreover, the average number of information bits received at user 2 from user 1  is identical to the average number of information bits that user 2  receives from the relay, i.e.,  $\bar{R}_{12}=\bar{R}_{r2}$. Similarly, we obtain that  $\bar{R}_{21}=\bar{R}_{r1}$ has to hold. Throughout this paper, the outage probability for each transmission direction is defined as the reduction in throughput compared to the ideal case when  $\boldsymbol{\gamma}(i)\in\mathcal{R}_1$ and $\boldsymbol{\ell}(i-1)\in\mathcal{L}_1$, for $\forall i$ hold \cite{NikolaMixed}. Mathematically, we write the flow outage probabilities as
\begin{IEEEeqnarray}{lll}\label{SysOut}
F^{\mathrm{out}}_{12} = 1- \frac{\bar{R}_{12}}{\bar{R}_{12}^{\max}} \quad \mathrm{and}\quad F^{\mathrm{out}}_{21} = 1- \frac{\bar{R}_{21}}{\bar{R}_{21}^{\max}}
\end{IEEEeqnarray}
where $\bar{R}_{12}^{\max}=\bar{R}_{21}^{\max}=R_0/2$. Furthermore, the sum throughput and the outage probability of the system are given by $\bar{R}^\mathrm{sum}=\bar{R}_{12}+\bar{R}_{21}$ and   $F^{\mathrm{out}}_{\mathrm{sys}} = \frac{F^{\mathrm{out}}_{12} + F^{\mathrm{out}}_{21}}{2}$, respectively.

Let  $s=(\ell_1,\ell_2),\,\,\ell_1=0,\dots,\ell_1^\mathrm{max},\,\ell_2=0,\dots,\ell_2^\mathrm{max}$ denote the states of the queues of the buffers at the relay. Moreover, for future reference, we define the transition probability $m_{s}^{s'} = \Pr\{s\to s'\}$, i.e.,  if the buffers are in state $s$ in the $(i-1)$-th time 
slot, with probability $m_{s}^{s'}$ the state of the queues is $s'$ in the $i$-th time slot. The queues of the buffers can be in $(1+\ell_1^{\max})(1+\ell_2^{\max})$ different states.

\begin{remk}\label{ZeroTransProb}
Since all nodes may transmit only one packet with a fixed transmission rate $R_0$ in each time slot, the number of packets in each queue may increase or decrease only by one packet. In other words, the transition probability $m_{(\ell_1,\ell_2)}^{(\ell_1',\ell_2')}$ is zero if $|\ell_1'\Minus \ell_1| \Gequal 2$ or $|\ell_2'\Minus \ell_2| \Gequal 2$. Moreover, due to the half-duplex constraint, the relay cannot transmit and receive at the same time. Therefore, the number of packets in one queue cannot increase if, in the same time slot, the number of the packets in the other queue decreases and vise versa, i.e., $m_{(\ell_1,\ell_2)}^{(\ell_1-1,\ell_2+1)}=0$ and $m_{(\ell_1,\ell_2)}^{(\ell_1+1,\ell_2-1)}=0$ hold.
\end{remk}

In order to analytically obtain the average throughput and the average delay, we first have to calculate the state occupancy probability of the Markov chain as the number of time slots tends to infinity, i.e., $\Pr\{s\Equal (\ell_1,\ell_2)\}$. To this end, all possible states are collected in one vector $\mathbf{s}$ as follows
\begin{IEEEeqnarray}{lll}  
		\mathbf{s}=\big[(0,0), \dots,(\ell_1^\mathrm{max},0),&(0,1), \dots,(\ell_1^\mathrm{max},1), \nonumber \\ 
		&(0,2), \dots,(\ell_1^\mathrm{max},\ell_2^\mathrm{max}) \big]^T, \quad
\end{IEEEeqnarray}
where $[\cdot]^T$ denotes the transpose operation. Moreover, $\mathbf{s}(n)$ refers to the state in the $n$-th element of vector $\mathbf{s}$. Furthermore,  vector $\Pr\{\mathbf{s}\}=[\Pr\{s\Equal(0,0)\},\dots,\Pr\{s\Equal(\ell_1^\mathrm{max},\ell_2^\mathrm{max})\}]^T$ contains all state occupancy probabilities. Let $\mathbf{M}$ denote the state transition matrix of the Markov chain where the entry in the $m$-th column and $n$-th row of $\mathbf{M}$ represents the transition probability from state $\mathbf{s}(m)$ to state $\mathbf{s}(n)$, i.e., $m_{\mathbf{s}(m)}^{\mathbf{s}(n)}$.

A state $m$ is accessible from state $n$ if the transition from state $n$ to state $m$ is possible with non-zero probability
in a finite number of steps \cite{StochGallager}. In this paper, we refer to the set of states which are accessible from the initial state $s=(0,0)$ as the reduced Markov chain. Note that it suffices to only  consider the reduced Markov chain for the performance analysis. Moreover, the state occupancy probability of a Markov chain can be obtained from the following three linear equations
\begin{IEEEeqnarray}{lll}\label{SteadyEqu}  
		\mathbf{M}\Pr\{\mathbf{s}\}=\Pr\{\mathbf{s}\},\quad
		\mathbf{1}^T\Pr\{\mathbf{s}\}=1,\quad
		\Pr\{\mathbf{s}\}\geq \mathbf{0}
\end{IEEEeqnarray}
where $\mathbf{1}$ and $\mathbf{0}$ denote $M$-dimensional vectors with all elements equal to one and zero, respectively, and $M$ is the number of states in the reduced Markov chain.

Using the transition and state occupancy probabilities, the throughputs of both information flows are obtained as 
\begin{IEEEeqnarray}{lll}\label{Throughput}  
		\bar{R}_{12} \hspace{-1mm}\Equal \hspace{-1mm}\sum_{\ell_1\Equal 1}^{\ell_1^\mathrm{max}} \hspace{-1mm} \sum_{\ell_2\Equal 1}^{\ell_2^\mathrm{max}} \hspace{-1mm} \left[ m_{(\ell_1,\ell_2)}^{(\ell_1\Minus 1,\ell_2)} \hspace{-1mm} \Add  m_{(\ell_1,\ell_2)}^{(\ell_1\Minus 1,\ell_2\Minus 1)} \hspace{-0.5mm} \right] \hspace{-1mm} \Pr\{s\Equal (\ell_1,\ell_2)\}  R_0  \IEEEyesnumber\IEEEyessubnumber \\
		\bar{R}_{21} \hspace{-1mm}\Equal \hspace{-1mm}\sum_{\ell_1\Equal 1}^{\ell_1^\mathrm{max}}\hspace{-1mm} \sum_{\ell_2\Equal 1}^{\ell_2^\mathrm{max}} \hspace{-1mm} \left[ m_{(\ell_1,\ell_2)}^{(\ell_1,\ell_2\Minus 1)} \hspace{-1mm} \Add m_{(\ell_1,\ell_2)}^{(\ell_1\Minus 1,\ell_2\Minus 1)} \hspace{-0.5mm} \right] \hspace{-1mm}\Pr\{s\Equal (\ell_1,\ell_2)\} R_0  \,\,\,\, \quad\IEEEyessubnumber
\end{IEEEeqnarray}
Moreover, let $T_j(i),\,\,j=1,2$, denote the waiting time that a packet transmitted from user $j$ in the $i$-th time slot  stays in buffer $B_j$ before it is transmitted to the respective user.   According to Little's Law \cite{Little}, the average delays of both information flows are obtained as 
\begin{IEEEeqnarray}{lll}\label{LittleLaw}  
		 \bar{T}_1 = \frac{\bar{Q}_1}{\bar{R}_{12}}\quad\mathrm{and}\quad\bar{T}_2 = \frac{\bar{Q}_1}{\bar{R}_{21}}
\end{IEEEeqnarray}
where 
\begin{IEEEeqnarray}{lll}\label{AveQ}  
		\bar{Q}_{1} \Equal \sum_{\ell_1\Equal 1}^{\ell_1^\mathrm{max}} \sum_{\ell_2\Equal 0}^{\ell_2^\mathrm{max}} \ell_1\Pr\{s\Equal (\ell_1,\ell_2)\}  \IEEEyesnumber\IEEEyessubnumber \\
		\bar{Q}_{2} \Equal \sum_{\ell_1\Equal 0}^{\ell_1^\mathrm{max}} \sum_{\ell_2\Equal 1}^{\ell_2^\mathrm{max}} \ell_2\Pr\{s\Equal (\ell_1,\ell_2)\}. \IEEEyessubnumber
\end{IEEEeqnarray}
Considering (\ref{SteadyEqu})-(\ref{AveQ}), the throughput and delay performances of any protocol with adaptive mode selection for the  bidirectional relaying network considered in this paper can be analytically evaluated once the transition probabilities of the states of the reduced Markov chain are determined.

Note that delay-constrained protocols with adaptive mode selection cannot surpass the sum throughput of the delay-unconstrained protocol given in \cite{ICCIEEE}. To also have a benchmark for the achievable minimum average delay, in the following, we provide a lower bound for the average delay such that no delay-constraint protocol with adaptive mode selection and  causal CSI information can achieve a lower average delay.

\begin{lem}\label{LemmMinDelay}
If adaptive mode selection based on causal CSI information is performed, the minimum achievable average delays for both information flows are given by
\begin{IEEEeqnarray}{lll}\label{MinDelay}  
		\bar{T}_1^{\min} &= \frac{1}{P_{\mathcal{R}_1}+P_{\mathcal{R}_2}+P_{\mathcal{R}_4}}, \,\,
		\bar{T}_2^{\min} &= \frac{1}{P_{\mathcal{R}_1}+P_{\mathcal{R}_2}+P_{\mathcal{R}_3}},\quad\,\,
\end{IEEEeqnarray}
where $P_{\mathcal{R}_m}=\Pr\{\boldsymbol{\gamma}(i) \in \mathcal{R}_m\}$. 
\end{lem}

\begin{figure*}[!t] 
\begin{IEEEeqnarray}{lll} \label{TransProb_Prot1Eq}
		 m_{(\ell_1,\ell_2)}^{(\ell_1',\ell_2')} &\hspace{-0.1cm}\Equal 0, \mathrm{if} \, |\ell_1'\Minus \ell_1| \Gequal 2 \vee |\ell_2'\Minus \ell_2| \Gequal 2 
		 \vee \hspace{-0.05cm} \{\ell_1'=\ell_1\Minus 1 \hspace{-0.05cm} \wedge \hspace{-0.05cm} \ell_2'=\ell_2\Add 1\} \hspace{-0.05cm} \vee \hspace{-0.05cm} \{\ell_1'=\ell_1\Add 1 \hspace{-0.05cm} \wedge \hspace{-0.05cm} \ell_2'=\ell_2\Minus 1\} \quad \IEEEyesnumber\IEEEyessubnumber \\
		m_{(\ell_1,\ell_2)}^{(\ell_1,\ell_2)} &\hspace{-0.1cm}\Equal \hspace{-0.1cm} \begin{cases}
		P_{\mathcal{R}_5}\Add P_{\mathcal{R}_3},\,&\mathrm{if} \,\ell_1 \Great \ell_1^{\mathrm{thr}}\,\wedge\,\ell_2 \Equal 0\\
		P_{\mathcal{R}_5}\Add P_{\mathcal{R}_4},\,&\mathrm{if} \,\ell_1 \Equal 0\,\wedge\,\ell_2 \Great \ell_2^{\mathrm{thr}}\\
		P_{\mathcal{R}_5},\,\,&\mathrm{otherwise} 
		\end{cases} \IEEEyessubnumber \\
		m_{(\ell_1,\ell_2)}^{(\ell_1\Add 1,\ell_2)} &\hspace{-0.1cm}\Equal \hspace{-0.1cm} \begin{cases}
		P_{\mathcal{R}_1}\Add P_{\mathcal{R}_2} \Add P_{\mathcal{R}_3},\, & \mathrm{if} \,\ell_2 \Great \ell_2^{\mathrm{thr}}\,\wedge\,\ell_1\Add \ell_2 \Less \ell_1^{\mathrm{thr}}\Add \ell_2^{\mathrm{thr}}\\
		P_{\mathcal{R}_2} \Add P_{\mathcal{R}_3},\,&\mathrm{if} \,\ell_2 \Lequal \ell_2^{\mathrm{thr}}\,\wedge\,\ell_1\Minus \ell_2 \Less \ell_1^{\mathrm{thr}}\Minus \ell_2^{\mathrm{thr}}\\
		\frac{P_{\mathcal{R}_2}}{2} \Add P_{\mathcal{R}_3},\,&\mathrm{if} \,\left\{\ell_2 \Less \ell_2^{\mathrm{thr}}\,\wedge\,\ell_1\Minus \ell_2 \Equal \ell_1^{\mathrm{thr}}\Minus \ell_2^{\mathrm{thr}}\right\}\,\vee\,\left\{ (\ell_1,\ell_2) \Equal (\ell_1^{\mathrm{thr}}, \ell_2^{\mathrm{thr}}) \Equal(0,0)\right\}\\
		\frac{P_{\mathcal{R}_2}\Add P_{\mathcal{R}_3}}{2},&\mathrm{if} \,\ell_1=0\,\wedge\,\ell_2=\ell_2^{\mathrm{thr}}\,\wedge\,\ell_1^{\mathrm{thr}}=0\\
		P_{\mathcal{R}_3},\,&\mathrm{if} \,\left\{\ell_1 \Less \ell_1^{\mathrm{thr}}\,\wedge\,\ell_1\Minus \ell_2 \Great \ell_1^{\mathrm{thr}}\Minus \ell_2^{\mathrm{thr}}\right\}
		\,\vee\,\left\{  (\ell_1,\ell_2) \Equal (\ell_1^{\mathrm{thr}},0) \right\}\\
		\frac{P_{\mathcal{R}_3}}{2},\,&\mathrm{if} \,\left\{\ell_1 \Equal \ell_1^{\mathrm{thr}}\,\wedge\,0\Less \ell_2 \Lequal \ell_2^{\mathrm{thr}} \,\wedge\,(\ell_1,\ell_2)\neq (0,
		\ell_2^{\mathrm{thr}}) \right\} \\ &\,\vee\,\left\{\ell_2 \Great \ell_2^{\mathrm{thr}}\,\wedge\,\ell_1\Add \ell_2 \Equal \ell_1^{\mathrm{thr}}\Add \ell_2^{\mathrm{thr}}  \, \wedge\, (\ell_1,\ell_2) \Nequal (0,\ell_1^{\mathrm{thr}}\Add \ell_2^{\mathrm{thr}})  \right\}\\
		\frac{P_{\mathcal{R}_1}\Add P_{\mathcal{R}_2} \Add P_{\mathcal{R}_3}}{2},\,&\mathrm{if} \, (\ell_1,\ell_2) \Equal (0,\ell_1^{\mathrm{thr}}\Add\ell_2^{\mathrm{thr}})\,\wedge\,\left\{\ell_1^{\mathrm{thr}}\Nequal 0 \,\vee \, \ell_2^{\mathrm{thr}}\Nequal 0 \right\}\\
		0,\,\,&\mathrm{otherwise} 
		\end{cases}  \IEEEyessubnumber \\
		m_{(\ell_1,\ell_2)}^{(\ell_1,\ell_2\Minus 1)} &\hspace{-0.1cm}\Equal \hspace{-0.1cm} \begin{cases}		
		P_{\mathcal{R}_1}\Add P_{\mathcal{R}_2} \Add P_{\mathcal{R}_3},\,&\mathrm{if} \,\ell_1 \Equal 0\,\wedge\, \ell_2 \Great \ell_1^{\mathrm{thr}}\Add \ell_2^{\mathrm{thr}}\\
		\frac{P_{\mathcal{R}_2}\Add P_{\mathcal{R}_3}}{2},&\mathrm{if} \,\ell_1=0\,\wedge\,\ell_2=\ell_2^{\mathrm{thr}}\,\wedge\,\ell_1^{\mathrm{thr}}=0\\
		P_{\mathcal{R}_3},\,&\mathrm{if} \,\left\{ \ell_1 \Great \ell_1^{\mathrm{thr}}\,\vee\, \ell_1\Add\ell_2 \Great \ell_1^{\mathrm{thr}} \Add \ell_2^{\mathrm{thr}} \right\} \, \wedge\,\ell_2\Nequal 0\\
		\frac{P_{\mathcal{R}_3}}{2},\,&\mathrm{if} \,\left\{\ell_1 \Equal \ell_1^{\mathrm{thr}}\,\wedge\,0\Less \ell_2 \Lequal \ell_2^{\mathrm{thr}} \right\}\,\vee \,\Big\{\ell_2 \Great \ell_2^{\mathrm{thr}}\,\wedge\,\ell_1\Add \ell_2 \Equal \ell_1^{\mathrm{thr}}\Add \ell_2^{\mathrm{thr}} \\ &\qquad\qquad\qquad\qquad\qquad\qquad\qquad\qquad\qquad  \wedge \, (\ell_1,\ell_2) \Nequal (0,\ell_1^{\mathrm{thr}}\Add \ell_2^{\mathrm{thr}}) \Big\}\\
		\frac{P_{\mathcal{R}_1}\Add P_{\mathcal{R}_2} \Add P_{\mathcal{R}_3}}{2},\,&\mathrm{if} \,(\ell_1,\ell_2) \Equal (0,\ell_1^{\mathrm{thr}}\Add \ell_2^{\mathrm{thr}})\,\wedge\,\left\{\ell_1^{\mathrm{thr}}\Nequal 0 \,\vee \, \ell_2^{\mathrm{thr}}\Nequal 0 \right\}\\
		0,\,\,&\mathrm{otherwise} 
		\end{cases}\quad  \IEEEyessubnumber \\
		m_{(\ell_1,\ell_2)}^{(\ell_1\Add 1,\ell_2\Add 1)} &\hspace{-0.1cm}\Equal \hspace{-0.1cm} \begin{cases}
		P_{\mathcal{R}_1},\,&\mathrm{if} \,\ell_1 \Lequal \ell_1^{\mathrm{thr}}\,\wedge\,\ell_2 \Lequal \ell_2^{\mathrm{thr}}\,\wedge\,\left\{(\ell_1,\ell_2)\Nequal (\ell_1^{\mathrm{thr}},\ell_2^{\mathrm{thr}}) \,\vee\,\ell_1^{\mathrm{thr}}\Equal 0\,\vee\,\ell_2^{\mathrm{thr}}\Equal 0\right\}\\
		\frac{P_{\mathcal{R}_1}}{2},\,&\mathrm{if} \,\ell_1 \Equal \ell_1^{\mathrm{thr}}\,\wedge\,\ell_2 \Equal \ell_2^{\mathrm{thr}}\,\wedge\,\ell_1^{\mathrm{thr}}\Nequal 0\,\wedge\,\ell_2^{\mathrm{thr}}\Nequal 0\\
		0,\,\,&\mathrm{otherwise} 
		\end{cases}  \IEEEyessubnumber \\
		m_{(\ell_1,\ell_2)}^{(\ell_1\Minus 1,\ell_2\Minus 1)} &\hspace{-0.1cm}\Equal \hspace{-0.1cm} \begin{cases}
		P_{\mathcal{R}_1}\Add P_{\mathcal{R}_2},\,&\mathrm{if} \,\ell_1 \Add \ell_2 \Gequal \ell_1^{\mathrm{thr}} \Add \ell_2^{\mathrm{thr}}\,\wedge\,\ell_1\Nequal 0 \,\wedge\, \ell_2\Nequal 0 \,\wedge\,(\ell_1,\ell_2)\Nequal (\ell_1^{\mathrm{thr}},\ell_2^{\mathrm{thr}})\\
		\frac{P_{\mathcal{R}_1}}{2}\Add P_{\mathcal{R}_2},\,\hspace{-0.4cm}&\mathrm{if} \,\ell_1 \Equal \ell_1^{\mathrm{thr}}\,\wedge\,\ell_2 \Equal \ell_2^{\mathrm{thr}}\,\wedge\,\ell_1^{\mathrm{thr}}\Nequal 0\,\wedge\,\ell_2^{\mathrm{thr}}\Nequal 0\\
		0,\,\,&\mathrm{otherwise} 
		\end{cases}  \hspace{-0.5cm}\IEEEyessubnumber
\end{IEEEeqnarray}
\vspace{-0.5cm}
\end{figure*}

\begin{IEEEproof}
In order to obtain the minimum possible average delay, a packet in the queue has to be retransmitted to the respective destination as soon as the relay-to-destination link can support the transmission at the chosen rate $R_0$. Thereby, the minimum delays of packets, $T_1^{\min}$ and $T_2^{\min}$,  have  geometric distributions with probability mass functions $f_{T_1^{\min}}(T_1^{\min}) = (1-p_2)^{T_1^{\min}-1}p_2,\,\,T_1^{\min}=1,2,\dots$ and $f_{T_2^{\min}}(T_2^{\min}) = (1-p_1)^{T_1^{\min}-1}p_1,\,\,T_2^{\min}=1,2,\dots$, respectively, where $p_1=\Pr\{\gamma_1(i)\geq \gamma_{\mathrm{thr}}\}$ and $p_2=\Pr\{\gamma_2(i)\geq \gamma_{\mathrm{thr}}\}$. Moreover, the means of the geometric random variables $T_1^{\min}$ and $T_2^{\min}$ are given by $\bar{T}_1^{\min}=\frac{1}{p_2}$ and $\bar{T}_2^{\min}=\frac{1}{p_1}$, respectively. This leads to (\ref{MinDelay}) and completes the proof.
\end{IEEEproof}

\begin{remk}
In this paper, we do not consider the case when the target average delays are below the limits in Lemma \ref{LemmMinDelay}. For such small average delay requirements, one can use the conventional bidirectional relaying protocols which do not perform adaptive mode selection \cite{TDBC,MABC,Tarokh}. Note that the conventional protocols with their fixed tranmission schedules cause an information loss when the user-to-relay transmission modes are active during outage events, which is not the case for the proposed protocols with adaptive mode selection. In fact, the lower average delays that the conventional protocols can achieve come at the expense of this information loss.
\end{remk}

\subsection{Main Results for the Proposed Protocol}

\iftoggle{paper}{
  
}{

We first provide the results for the delay-efficient mode, and then the results for throughput-efficient mode.

\vspace{0.3cm}
\noindent
\textit{1) Delay-Efficient Mode}
\vspace{0.1cm}  
}

In the following, we present the transition probabilities of the proposed protocol in the delay-efficient mode. 

\iftoggle{paper}{

\begin{Prop}\label{TransProb_Prot1}
The transition probabilities $m_{s}^{s'}$ of the states of the Markov chain for the proposed protocol in the delay-efficient mode are given by (\ref{TransProb_Prot1Eq}) on the top of the next page, where the values of $m_{(\ell_1,\ell_2)}^{(\ell_1,\ell_2\Add 1)}$ and $m_{(\ell_1,\ell_2)}^{(\ell_1\Minus 1,\ell_2)}$ are identical to the values of $m_{(\ell_1,\ell_2)}^{(\ell_1\Add 1,\ell_2)} $ and $m_{(\ell_1,\ell_2)}^{(\ell_1,\ell_2\Minus 1)}$, respectively, after switching the roles of user 1 and user 2.
\end{Prop}
  
}{

\begin{Prop}\label{TransProb_Prot1}
The transition probabilities $m_{s}^{s'}$ of the states of the Markov chain for the proposed protocol in the delay-efficient mode are given by (\ref{TransProb_Prot1Eq}) on the top of this page, where the values of $m_{(\ell_1,\ell_2)}^{(\ell_1,\ell_2\Add 1)}$ and $m_{(\ell_1,\ell_2)}^{(\ell_1\Minus 1,\ell_2)}$ are identical to the values of $m_{(\ell_1,\ell_2)}^{(\ell_1\Add 1,\ell_2)} $ and $m_{(\ell_1,\ell_2)}^{(\ell_1,\ell_2\Minus 1)}$, respectively, after switching the roles of user 1 and user 2.
\end{Prop}
  
}

\iftoggle{paper}{

\begin{IEEEproof}
For the transition probabilities in (\ref{TransProb_Prot1Eq}a), we refer to Remark \ref{ZeroTransProb}.  For the remaining transition probabilities, we use the following partitioning of the possible SNR regions
\begin{IEEEeqnarray}{lll}\label{Partition}  
	m_{s}^{s'} \hspace{-1mm}\Equal \hspace{-1mm} \sum_{m=1}^{5} \hspace{-0.5mm} P_{\mathcal{R}_m} \Pr \left\{\boldsymbol{\ell}(i)=s'|\boldsymbol{\gamma}(i)\in\mathcal{R}_m \wedge \boldsymbol{\ell}(i-1)=s \right\}	 \quad\,\,
\end{IEEEeqnarray}
In particular, by applying this partitioning for the different cases distinguished in (\ref{TransProb_Prot1Eq}), we obtain the given values of $m_{s}^{s'}$. Due to the space limitation, we have to refer to \cite[Appendix A]{GlobeCom2014Arxiv} for a detailed proof. 
\end{IEEEproof}
  
}{

\begin{IEEEproof}
Please refer to Appendix \ref{AppTransProt1}.
\end{IEEEproof}
  
}

For the proposed protocol in the delay-efficient mode, states $s=(\ell_1,\ell_2)$ for which $\ell_1>\ell_1^{\mathrm{thr}}+1$ or $\ell_2>\ell_2^{\mathrm{thr}}+1$ hold, are not accessible from the initial state $s=(0,0)$. Therefore, it is sufficient to only consider the states of the queues corresponding to $\ell_1\leq \ell_1^{\mathrm{thr}}+1$ and $\ell_2\leq \ell_2^{\mathrm{thr}}+1$ to calculate the state occupancy probabilities based on (\ref{SteadyEqu}). Using the transition probabilities given in Proposition \ref{TransProb_Prot1} and the state occupancy probabilities, the average throughput and the average delay of the proposed  protocol in the delay-efficient mode can be obtained from (\ref{Throughput}) and (\ref{LittleLaw}), respectively.

\begin{Prop}\label{PropMinDelayProt1}
The minimum target average delays that the proposed protocol in the delay-efficient mode can support are 
\begin{IEEEeqnarray}{lll} \label{MinDelayProt1}  
		\bar{T}_1^{\mathrm{d}}=\frac{1}{P_{\mathcal{R}_1}+P_{\mathcal{R}_2}+P_{\mathcal{R}_4}},\,\,  
		\bar{T}_2^{\mathrm{d}}= \frac{1}{P_{\mathcal{R}_1}+P_{\mathcal{R}_2}+P_{\mathcal{R}_3}}, \quad
\end{IEEEeqnarray}
and the achievable average throughputs with the above  average delay constraints are given by 
\begin{IEEEeqnarray}{lll} \label{MinThrProt1}  
		\bar{R}_{12}= \frac{a+b}{1+a+b+c}(P_{\mathcal{R}_1}+P_{\mathcal{R}_2}+P_{\mathcal{R}_4})R_0 \IEEEyesnumber\IEEEyessubnumber\\ 
		\bar{R}_{21}= \frac{a+c}{1+a+b+c}(P_{\mathcal{R}_1}+P_{\mathcal{R}_2}+P_{\mathcal{R}_3})R_0,  \IEEEyessubnumber
\end{IEEEeqnarray}
where $a$, $b$, and $c$ are given by
\begin{IEEEeqnarray}{lll}\label{abc} 
		a = \frac{P_{\mathcal{R}_1}}{1-P_{\mathcal{R}_5}} \IEEEyesnumber\IEEEyessubnumber\\
		b = \frac{1}{1-P_{\mathcal{R}_3}-P_{\mathcal{R}_5}}\left[P_{\mathcal{R}_3}+\frac{P_{\mathcal{R}_2}}{2}+\frac{P_{\mathcal{R}_1}P_{\mathcal{R}_3}}{1-P_{\mathcal{R}_5}} \right] \IEEEyessubnumber\\
		c = \frac{1}{1-P_{\mathcal{R}_4}-P_{\mathcal{R}_5}}\left[P_{\mathcal{R}_4}+\frac{P_{\mathcal{R}_2}}{2}+\frac{P_{\mathcal{R}_1}P_{\mathcal{R}_4}}{1-P_{\mathcal{R}_5}} \right]. \IEEEyessubnumber
\end{IEEEeqnarray}
\end{Prop}

\iftoggle{paper}{

\begin{IEEEproof}
The lowest values for the average delays for both information flows are obtained when $\ell_1^{\mathrm{thr}}\Equal \ell_2^{\mathrm{thr}}\Equal0$. We obtain the results in Proposition \ref{PropMinDelayProt1} by substituting the transition probabilities in Proposition \ref{TransProb_Prot1} for $\ell_1^{\mathrm{thr}}\Equal\ell_2^{\mathrm{thr}}\Equal 0$ into (\ref{SteadyEqu}), (\ref{Throughput}), and (\ref{LittleLaw}).  We refer to \cite[Appendix B]{GlobeCom2014Arxiv} for a more detailed proof. 
\end{IEEEproof}
  
}{

\begin{IEEEproof}
Please refer to Appendix \ref{AppProt1MinDel}.
\end{IEEEproof}
  
}

\begin{remk}
Note that the minimum average delays given in Proposition \ref{PropMinDelayProt1} are indeed the minimum possible average delays that can be achieved by any adaptive mode selection protocol, cf.  (\ref{MinDelay}) in Lemma \ref{LemmMinDelay}. Therefore, the proposed protocol in the delay-efficient mode can achieve all  possible average delays which makes this protocol attractive for strictly delay-constrained applications. 
\end{remk}

\iftoggle{paper}{
  
}{

\vspace{0.3cm}
\noindent
\textit{2) Throughput-Efficient Mode}
\vspace{0.1cm}

In order to avoid repetition, we do not provide the expressions for the transition probabilities of the proposed protocol in the throughput-efficient mode for the general case. However, we provide the state occupancy probabilities for the minimum possible average delay in the following proposition.  Based on this result, the minimum average delays and the respective average throughputs can be calculated.

\begin{Prop}\label{PropMinDelayProt2Modified}
The state occupancy probabilities of the proposed protocol in the throughput-efficient mode and the minimum target average delays, i.e., $\ell_1^{\mathrm{thr}}=\ell_2^{\mathrm{thr}}=0$, are given by
\begin{IEEEeqnarray}{lll} \label{StateOccProbProt2-TE}
	\Pr\{s(\ell_1,\ell_2)\} = f(\ell_1,\ell_2)x + g(\ell_1,\ell_2)y,
	\end{IEEEeqnarray}
where $x$ and $y$ are defined as
\begin{IEEEeqnarray}{lll}
	x =& \frac{1}{\sum_{\ell_1=1}^{\ell_1^{\max}}\sum_{\ell_2=1}^{\ell_2^{\max}} f(\ell_1,\ell_2)+z g(\ell_1,\ell_2)},\,\, y = zx,
	\end{IEEEeqnarray}
and
\begin{IEEEeqnarray}{lll}
	z \Equal & \bigg[ (1\Minus P_{\mathcal{R}_5})f(1,0) \Minus (P_{\mathcal{R}_2}\Add P_{\mathcal{R}_4})f(2,0) \Minus (P_{\mathcal{R}_1}\Add P_{\mathcal{R}_2}) \nonumber \\
	&f(2,1) \Minus \left(\frac{P_{\mathcal{R}_2}}{2}\Add P_{\mathcal{R}_3}\right)f(0,0)\Add P_{\mathcal{R}_3}f(1,1) \bigg ] \nonumber \\ & / \bigg[ \left(\frac{P_{\mathcal{R}_2}}{2}\Add P_{\mathcal{R}_3}\right)g(0,0)\Add P_{\mathcal{R}_3}g(1,1)\bigg].
	\end{IEEEeqnarray}
Furthermore, $f(\ell_1,\ell_2)=f(0,1)=0,\,\,\ell_2>2$ and $g(\ell_1,\ell_2)=g(1,0)=0,\,\,\ell_1>2$ hold. The non-zero values of $f(\ell_1,\ell_2)$ are given by
\begin{IEEEeqnarray}{lll} 
	f(\ell_1,0)\Equal &\frac{1}{P_{\mathcal{R}_3}\Add\frac{P_{\mathcal{R}_1}P_{\mathcal{R}_3}}{1\Minus P_{\mathcal{R}_5}}}\bigg[ (1\Minus P_{\mathcal{R}_5}) f(\ell_1\Add 1,0)  \nonumber \\  
	& \hspace{-1cm}\Minus (P_{\mathcal{R}_2}\Add P_{\mathcal{R}_4}) f(\ell_1\Add 2,0)    \Minus \left(P_{\mathcal{R}_1}\Add P_{\mathcal{R}_2}+\frac{P_{\mathcal{R}_3} P_{\mathcal{R}_4}}{1\Minus P_{\mathcal{R}_5}}\right)  \nonumber \\   
	& \hspace{-1cm} f(\ell_1\Add 2,1) \bigg],  \qquad \qquad \qquad \mathrm{if}\,\,\ell_1\Equal 1,\cdots,\ell_1^{\mathrm{\max}}\Minus 2 \IEEEyesnumber\IEEEyessubnumber\\
	f(\ell_1,1) \Equal & \frac{1}{1\Minus P_{\mathcal{R}_5}} \left[ P_{\mathcal{R}_4} f(\ell_1\Add 1,1) + P_{\mathcal{R}_1}f(\ell_1\Minus  1,0) \right],   \nonumber \\ &\qquad \qquad \qquad \qquad \,\,\,\,\, \mathrm{if}\,\,\ell_1\Equal 2,\cdots,\ell_1^{\mathrm{\max}}\Minus  1 \IEEEyessubnumber
	\end{IEEEeqnarray}
where $f(\ell_1^{\mathrm{\max}},0)=1$, $f(\ell_1^{\mathrm{\max}},1)=\frac{P_{\mathcal{R}_1}(1-P_{\mathcal{R}_3}-P_{\mathcal{R}_5})}{P_{\mathcal{R}_3}(1+P_{\mathcal{R}_1}-P_{\mathcal{R}_5})}$, $f(\ell_1^{\mathrm{\max}}-1,0)=\frac{1-P_{\mathcal{R}_5}}{P_{\mathcal{R}_1}}f(\ell_1^{\mathrm{\max}},1)$, and $f(\ell_1^{\mathrm{\max}}-2,0)=\frac{1}{P_{\mathcal{R}_3}+\frac{P_{\mathcal{R}_1}P_{\mathcal{R}_3}}{1-P_{\mathcal{R}_5}}}\Big[ (1-P_{\mathcal{R}_5}) f(\ell_1^{\mathrm{\max}}-1,0) -(P_{\mathcal{R}_1}+P_{\mathcal{R}_2}+P_{\mathcal{R}_4})  -\left(P_{\mathcal{R}_1}+P_{\mathcal{R}_2}+\frac{P_{\mathcal{R}_3} P_{\mathcal{R}_4}}{1-P_{\mathcal{R}_5}}\right)f(\ell_1^{\mathrm{\max}},1) \Big]$. Similarly, the non-zero values of $g(\ell_1,\ell_2)$ are given by
\begin{IEEEeqnarray}{lll} 
	g(0,\ell_2)\Equal &\frac{1}{P_{\mathcal{R}_4}\Add \frac{P_{\mathcal{R}_1}P_{\mathcal{R}_4}}{1\Minus P_{\mathcal{R}_5}}}\bigg[ (1\Minus P_{\mathcal{R}_5}) g(0,\ell_2\Add 1) \nonumber \\  
	&\hspace{-1cm} \Minus (P_{\mathcal{R}_2}+P_{\mathcal{R}_3})  g(0,\ell_2\Add 2) \Minus \left(P_{\mathcal{R}_1}+P_{\mathcal{R}_2}\Add \frac{P_{\mathcal{R}_3} P_{\mathcal{R}_4}}{1\Minus P_{\mathcal{R}_5}}\right) \nonumber \\    
& \hspace{-1cm}	g(1,\ell_2\Add 2) \bigg], \qquad \qquad \qquad \mathrm{if}\,\, \ell_2\Equal 1,\cdots,\ell_2^{\mathrm{\max}}\Minus 2 \IEEEyesnumber\IEEEyessubnumber\\
	g(1,\ell_2) \Equal & \frac{1}{1\Minus P_{\mathcal{R}_5}} \left[ P_{\mathcal{R}_3} g(1,\ell_2\Add 1) \Add P_{\mathcal{R}_1}g(0,\ell_2\Minus 1) \right],\nonumber \\ &\qquad \qquad \qquad \qquad \,\,\,\,\, \mathrm{if}\,\, \ell_2\Equal 2,\cdots,\ell_2^{\mathrm{\max}}\Minus 1 \IEEEyessubnumber
	\end{IEEEeqnarray}
where $g(0,\ell_2^{\mathrm{\max}})\Equal 1$, $g(1,\ell_2^{\mathrm{\max}})=\frac{P_{\mathcal{R}_1}(1-P_{\mathcal{R}_4}-P_{\mathcal{R}_5})}{P_{\mathcal{R}_4}(1+P_{\mathcal{R}_1}-P_{\mathcal{R}_5})}$, $g(0,\ell_2^{\mathrm{\max}}-1)=\frac{1-P_{\mathcal{R}_5}}{P_{\mathcal{R}_1}}g(1,\ell_2^{\mathrm{\max}})$, and $g(0,\ell_2^{\mathrm{\max}}-2)=\frac{1}{P_{\mathcal{R}_4}+\frac{P_{\mathcal{R}_1}P_{\mathcal{R}_4}}{1-P_{\mathcal{R}_5}}}\Big[ (1-P_{\mathcal{R}_5}) g(0,\ell_2^{\mathrm{\max}}-1) -(P_{\mathcal{R}_1}+P_{\mathcal{R}_2}+P_{\mathcal{R}_3}) -\left(P_{\mathcal{R}_1}+P_{\mathcal{R}_2}+\frac{P_{\mathcal{R}_3} P_{\mathcal{R}_4}}{1-P_{\mathcal{R}_5}}\right)g(1,\ell_2^{\mathrm{\max}}) \Big]$. Moreover, $f(0,0)$, $g(0,0)$, $f(1,1)$, and $g(1,1)$ are given by
\begin{IEEEeqnarray}{lll}
	f(0,0) &= \frac{1-P_{\mathcal{R}_5}}{(1-P_{\mathcal{R}_5})^2-P_{\mathcal{R}_1}(P_{\mathcal{R}_1}+P_{\mathcal{R}_2})} \nonumber \\
	&\Big[ (P_{\mathcal{R}_2}\Add P_{\mathcal{R}_4})f(1,0) \Add  \frac{P_{\mathcal{R}_4}(P_{\mathcal{R}_1}  \Add P_{\mathcal{R}_2})}{1\Minus P_{\mathcal{R}_5}}f(2,1)\Big] \qquad \IEEEyesnumber\IEEEyessubnumber\\
	g(0,0) &= \frac{1-P_{\mathcal{R}_5}}{(1-P_{\mathcal{R}_5})^2-P_{\mathcal{R}_1}(P_{\mathcal{R}_1}+P_{\mathcal{R}_2})} \nonumber \\ 
	& \Big[ (P_{\mathcal{R}_2} \Add  P_{\mathcal{R}_3})g(0,1)\Add  \frac{P_{\mathcal{R}_3}(P_{\mathcal{R}_1} \Add  P_{\mathcal{R}_2})}{1\Minus P_{\mathcal{R}_5}}g(1,2)\Big] \IEEEyessubnumber\\
	f(1,1)&= \frac{1}{1-P_{\mathcal{R}_5}}\Big[ P_{\mathcal{R}_1} f(0,0) +P_{\mathcal{R}_4}f(2,1) \Big] \IEEEyessubnumber\\
	g(1,1)&= \frac{1}{1-P_{\mathcal{R}_5}}\Big[ P_{\mathcal{R}_1} g(0,0) +P_{\mathcal{R}_3}g(1,2) \Big], \IEEEyessubnumber
	\end{IEEEeqnarray}
respectively. 
\end{Prop}

\begin{IEEEproof}
Please refer to Appendix \ref{AppProt2Trans}.
\end{IEEEproof}
 
Note that the minimum average delays for the proposed protocol in the throughput-efficient mode are higher than those in the delay-efficient mode.

}

\subsection{High SNR Analysis of the Proposed Protocol}

\iftoggle{paper}{

Next, we investigate the performance of the proposed protocol in the high SNR regime for Rayleigh fading. For Rayleigh fading, the probability density functions (pdfs) of $\gamma_1(i)$ and $\gamma_2(i)$ are given by $f_{\gamma_1}(\gamma_1)=\frac{1}{\Omega_1\gamma}e^{-\frac{\gamma_1}{\Omega_1\gamma}}$ and $f_{\gamma_2}(\gamma_2)=\frac{1}{\Omega_2\gamma}e^{-\frac{\gamma_2}{\Omega_2\gamma}}$, respectively.   For future reference, we define $f(x)=o(g(x))$ if $\underset{x\to 0}{\lim} \frac{f(x)}{g(x)} = 0$.  
  
}{

Next, we investigate the performance of the proposed protocol in the high SNR regime for Rayleigh fading. For Rayleigh fading, the probability density functions (pdfs) of $\gamma_1(i)$ and $\gamma_2(i)$ are given by $f_{\gamma_1}(\gamma_1)=\frac{1}{\Omega_1\gamma}e^{-\frac{\gamma_1}{\Omega_1\gamma}}$ and $f_{\gamma_2}(\gamma_2)=\frac{1}{\Omega_2\gamma}e^{-\frac{\gamma_2}{\Omega_2\gamma}}$, respectively.   For high SNR analysis, we present the asymptotic values of $P_{\mathcal{R}_m},\,\,m=1,\dots,5$ in the high SNR regime, i.e., $\gamma\to\infty$, as
\begin{IEEEeqnarray}{lll}\label{P_Ray}
P_{\mathcal{R}_1} \hspace{-1mm} \Equal  1-\frac{(\Omega_1+\Omega_2)\gamma_{\mathrm{thr}}}{\Omega_1\Omega_2} \cdot  \frac{1}{\gamma}+o\left(\frac{1}{\gamma}\right) \qquad \IEEEyesnumber\IEEEyessubnumber \\
P_{\mathcal{R}_2} \hspace{-1mm} \Equal \frac{1}{\Omega_1\Omega_2} \hspace{-0.5mm} \left[ \hspace{-0.5mm} 2\gamma_{\mathrm{thr}}^2 \hspace{-0.5mm} \Add \hspace{-0.5mm} \frac{(\gamma_{\mathrm{thr}}^{\mathrm{sum}})^2}{2} \Minus \hspace{-0.5mm} 2\gamma_{\mathrm{thr}}\gamma_{\mathrm{thr}}^{\mathrm{sum}} \hspace{-0.5mm}\right]\hspace{-1mm}\cdot  \hspace{-1mm} \frac{1}{\gamma^2} \hspace{-0.5mm}\Add \hspace{-0.5mm} o\left(\hspace{-0.3mm}\frac{1}{\gamma^2}\hspace{-0.3mm}\right) \qquad \IEEEyessubnumber \\
P_{\mathcal{R}_3} \hspace{-1mm} \Equal \frac{\gamma_{\mathrm{thr}}}{\Omega_2} \cdot  \frac{1}{\gamma}+o\left(\frac{1}{\gamma}\right)  \IEEEyessubnumber \\
P_{\mathcal{R}_4} \hspace{-1mm}\Equal \frac{\gamma_{\mathrm{thr}}}{\Omega_1} \cdot  \frac{1}{\gamma}+o\left(\frac{1}{\gamma}\right)  \IEEEyessubnumber \\
P_{\mathcal{R}_5} \hspace{-1mm}\Equal \frac{\gamma^2_{\mathrm{thr}}}{\Omega_1\Omega_2} \cdot  \frac{1}{\gamma^2}+o\left(\frac{1}{\gamma^2}\right)  \IEEEyessubnumber
\end{IEEEeqnarray}
where we used the Taylor series $e^{x}=1+x+o\left(x\right),\,\,x\to 0$ for derivation of (\ref{P_Ray}) where $f(x)=o(g(x))$ if $\underset{x\to 0}{\lim} \frac{f(x)}{g(x)} = 0$. 
  
}

\iftoggle{paper}{
  
}{

\vspace{0.3cm}
\noindent
\textit{1) Delay-Efficient Mode}
\vspace{0.1cm}  
}

\begin{Prop}\label{PropHighSNRProt1}
In the high SNR regime, i.e., $\gamma\to\infty$, and for Rayleigh fading, the sum throughput and the system outage probability of the proposed  protocol in the delay-efficient mode and the minimum target average delays in Proposition \ref{PropMinDelayProt1} are given by
\begin{IEEEeqnarray}{lll}\label{HighThrOutPro1}
\bar{R}_{\mathrm{sum}} = R_0,\,\,
     F^{\mathrm{out}}_{\mathrm{sys}}  = \frac{(\Omega_1+\Omega_2)\gamma_{\mathrm{thr}}}{\Omega_1\Omega_2} \cdot \frac{1}{\gamma} + o\left(\frac{1}{\gamma}\right), 
\end{IEEEeqnarray}
where the target average delays approach one, i.e., $(\bar{T}_1^{\mathrm{d}},\bar{T}_2^{\mathrm{d}})\to (1,1)$ holds. Moreover, the individual throughputs and the outage probabilities of the users are given by
\begin{IEEEeqnarray}{ccc}\label{ThrHighIndPro1} 
\bar{R}_{12} = \bar{R}_{21} =  \frac{R_0}{2} \vspace{-0.3cm}
\end{IEEEeqnarray}
\begin{IEEEeqnarray}{ccc}\label{FoutHighIndPro1}
      F^{\mathrm{out}}_{12}  = \frac{(\Omega_1+3\Omega_2)\gamma_{\mathrm{thr}}}{2\Omega_1\Omega_2} \cdot \frac{1}{\gamma} + o\left(\frac{1}{\gamma}\right) \IEEEyesnumber\IEEEyessubnumber \\
       F^{\mathrm{out}}_{21}  = \frac{(3\Omega_1+\Omega_2)\gamma_{\mathrm{thr}}}{2\Omega_1\Omega_2} \cdot \frac{1}{\gamma} + o\left(\frac{1}{\gamma}\right), \IEEEyessubnumber
\end{IEEEeqnarray}
respectively.
\end{Prop}

\iftoggle{paper}{
  
  \begin{IEEEproof}
For $\gamma\to\infty$, exploiting the definitions of $\mathcal{R}_m$ and $P_{\mathcal{R}_m}$, we obtain $P_{\mathcal{R}_1}\to 1$, $P_{\mathcal{R}_m}\to 0,\,\,m=2,3,4,5$, $a\to 1$, and $b,c\to 0$. Hence, we obtain $\bar{R}_{12}\to \frac{R_0}{2}$ and $\bar{R}_{12}\to \frac{R_0}{2}$ from (\ref{MinThrProt1}), and consequently $\bar{R}_{\mathrm{sum}} \to R_0$. In a similar manner, we obtain $(\bar{T}_1^{\mathrm{d}},\bar{T}_2^{\mathrm{d}})\to (1,1)$ from (\ref{MinDelayProt1}). For derivation of the outage probabilities, we substitute the first order approximations of $P_{\mathcal{R}_m},\,\,m=1,\dots,5$ for Rayleigh fading, as given in \cite[eq. (23)]{GlobeCom2014Arxiv}, into (\ref{SysOut}) and (\ref{MinThrProt1}), and simplify the results using the Taylor series $\frac{1}{1+x}=1-x+o(x)$ for $x\to 0$ to obtain (\ref{FoutHighIndPro1}), and consequently the system outage probability in (\ref{HighThrOutPro1}). This completes the proof.
\end{IEEEproof}

}{

  \begin{IEEEproof}
Please refer to Appendix \ref{AppProt1High}.
\end{IEEEproof}

}

\begin{remk}
In the high SNR regime, the average sum throughput of the proposed protocol in the delay-efficient mode approaches the upper bound given in \cite{ICCIEEE} even for the minimum average delay of one time slot, cf. (\ref{HighThrOutPro1}). For the system outage probability, the same diversity order of one is obtained as for the delay-unconstrained protocol in \cite{ICCIEEE}. However, the system outage probabilities of the delay-unconstrained protocol and the proposed protocol in the delay-efficient mode with a target average delay of one time slot have an SNR gap of
\begin{IEEEeqnarray}{ccc}\label{SNRgap}
\text{SNR}_\mathrm{gap} = 10\log_{10}\left(1+\frac{\Omega_{\min}}{\Omega_{\max}}\right)\leq 3 \,\,\text{dB},
\end{IEEEeqnarray}
where $\Omega_{\min}=\min\{\Omega_1,\Omega_2\}$ and $\Omega_{\max}=\max\{\Omega_1,\Omega_2\}$. The expression in (\ref{SNRgap}) is obtained by comparing the required SNR for a given outage probability for the proposed protocol in (\ref{HighThrOutPro1}) and the protocol in \cite[eq. (9)]{ICCIEEE}.
\end{remk}

\begin{remk}
For the high SNR performance analysis in Proposition \ref{PropHighSNRProt1}, we assumed the minimum possible target average delays given in Proposition \ref{PropMinDelayProt1}.  We note that for larger permissible delays, the SNR gap is lower than that in (\ref{SNRgap}).
\end{remk}

\iftoggle{paper}{
  
}{

\vspace{0.3cm}
\noindent
\textit{2) Throughput-Efficient Mode}
\vspace{0.1cm}  

\begin{Prop}\label{PropHighSNRProt2Mod}
In the high SNR regime, i.e., $\gamma\to\infty$, for Rayleigh fading and symmetric channels, i.e., $\Omega_1=\Omega_2 \triangleq \Omega_{\mathrm{eq}}$, the sum throughput and the system outage probability of the proposed protocol in the throughput-efficient mode and the minimum target average delays, i.e., $\ell_1^{\mathrm{thr}}=\ell_2^{\mathrm{thr}}=0$, are given by
\begin{IEEEeqnarray}{lll}\label{HighThrOutPro2Modified}
\bar{R}_{\mathrm{sum}} = R_0 \quad \mathrm{and} \quad
     F^{\mathrm{out}}_{\mathrm{sys}}  = \frac{\gamma_{\mathrm{thr}}}{\Omega_{\mathrm{eq}}} \cdot \frac{1}{\gamma} + o\left(\frac{1}{\gamma}\right), 
\end{IEEEeqnarray}
where the resulting average delays are functions of the maximum lengths of the buffers and are given by
\begin{IEEEeqnarray}{lll}\label{HighDelayPro2Modified}
\bar{T}_1 \Equal \frac{\ell_1^{\max}.\ell_1^{\max} \Add \ell_2^{\max}\Minus 1}{\ell_1^{\max}\Add\ell_2^{\max}\Minus 1}, \,\, \bar{T}_2 \Equal \frac{\ell_2^{\max}.\ell_2^{\max}\Add \ell_1^{\max} \Minus 1}{\ell_1^{\max}\Add\ell_2^{\max} \Minus 1}.\,\,\quad
\end{IEEEeqnarray}
Moreover, the individual throughputs are given by $\bar{R}_{12}=\bar{R}_{12}=\frac{R_0}{2}$ and the individual outage probabilities are obtained as $F^{\mathrm{out}}_{12}=F^{\mathrm{out}}_{21}=F^{\mathrm{out}}_{\mathrm{sys}}$.
\end{Prop}

\begin{IEEEproof}
Please refer to Appendix \ref{AppProt2High}.
\end{IEEEproof}

\begin{remk} 
Note that the asymptotic system outage probability of the proposed protocol in the throughput-efficient mode converges to the lower bound given in \cite{ICCIEEE} for the delay-unconstrained protocol, i.e., the SNR gap vanishes.
\end{remk}

\begin{remk} 
For  asymmetric channels, for instance, $\Omega_1 > \Omega_2$, the proposed protocol in the throughput-efficient mode selects the user 1-to-relay and relay-to-user 1 transmission modes more frequently which increases $\bar{T}_1$ and decreases $\bar{T}_2$. Thus, for asymmetric channels, the proposed protocol in the throughput-efficient mode cannot efficiently limit the average delay of one of the information flows. For such cases, sufficiently small values for the maximum lengths of the buffers have to be chosen to limit the delay.
\end{remk}

}

\section{Numerical Results}

In this section, the performances of the proposed delay-constrained protocol is evaluated for 
Rayleigh fading. We consider the sum throughputs and the outage probabilities of the proposed protocol for the minimum possible average delays, i.e., for $\ell_1^{\mathrm{thr}}=\ell_2^{\mathrm{thr}}=0$. Moreover, the results are compared with the sum throughput and the system outage probability of the delay-unconstrained protocol in \cite{ICCIEEE}. We are interested in these comparisons, since the throughputs and the system outage probabilities of the proposed protocol with any valid values of the design variables, i.e., $\ell_j^{\mathrm{thr}}$, $j\Equal 1,2$, fall between the throughputs and the system outage probabilities of the aforementioned cases, respectively. Furthermore, we assume  $\ell_1^{\max}\Equal\ell_2^{\max}\Equal 10$, $R_0\Equal 1$, and consider both a symmetric channel, $\Omega_1\Equal \Omega_2\Equal 1$, and an asymmetric channel, $\Omega_1\Equal 0.25,\Omega_2\Equal 1$. The curves depicted in this section are obtained by analytically evaluating the proposed protocols\footnote{Note that all results have been verified by simulation. However, for clarity of presentation, the simulation results are not included in the figures.}.

As a benchmark scheme, we adopt the MABC protocol which, for high SNRs, has a superior performance  compared to the TDBC  and  traditional two-way relaying protocols. Recall that in the conventional MABC protocol, the relay receives information from both users in one time slot and forwards it to the respective users in the following time slot \cite{MABC}. For a fair comparison, we also consider a MABC protocol where the relay has a buffer and receives and stores information for $N/2$ consecutive  time slots and forwards them to the respective users in the remaining time slots. Note that the conventional MABC protocol has a average delay of one time slot for both information flows while the version employing buffers and $N\to\infty$ is delay-unlimited. Furthermore, for clarity of presentation, we only show results for the two considered MABC protocols for the  asymmetric channel.

\begin{figure}
\centering
\resizebox{0.9\linewidth}{!}{
\psfragfig{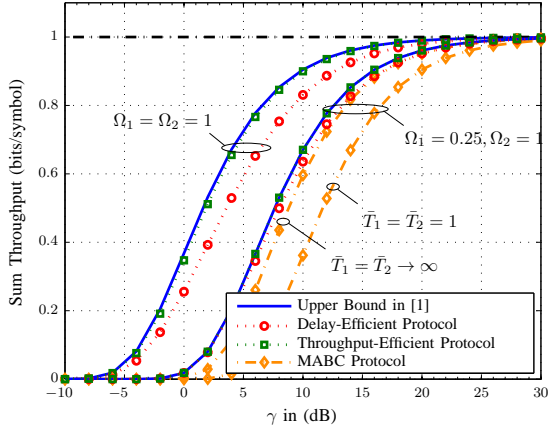}}
\vspace{-0.5cm}
\caption{Sum throughput vs. transmit SNR, $\gamma$, in (dB)  for $\ell_1^{\max}=\ell_2^{\max}=10$, $R_0=1$, and the minimum possible average delays. }
\label{FigRate}
\vspace{-0.5cm}
\end{figure}

In Fig. \ref{FigRate}, the average sum throughput, $\bar{R}^{\mathrm{sum}}$, is depicted versus the transmit SNR, $\gamma$, in (dB). We observe that the sum throughputs of all considered protocols converge to $R_0$ in the high SNR regime. Moreover, the average sum throughput of the proposed protocol in the throughput-efficient mode is higher than the average sum throughput of the proposed protocol in the delay-efficient mode.  Note that this gain is obtained at the cost of higher average delay of the protocol. We can also conclude that the SNR gap between the upper bound on the sum throughput given in \cite{ICCIEEE} and the sum throughput of the proposed protocols is smaller for asymmetric channels than for symmetric channels. Furthermore, there is a considerable performance gain compared to the MABC protocol with one time slot delay. Moreover, the proposed protocol with the minimum possible average delays even outperforms the MABC protocol with unlimited delay.

\begin{figure}
\centering
\resizebox{0.9\linewidth}{!}{
\psfragfig{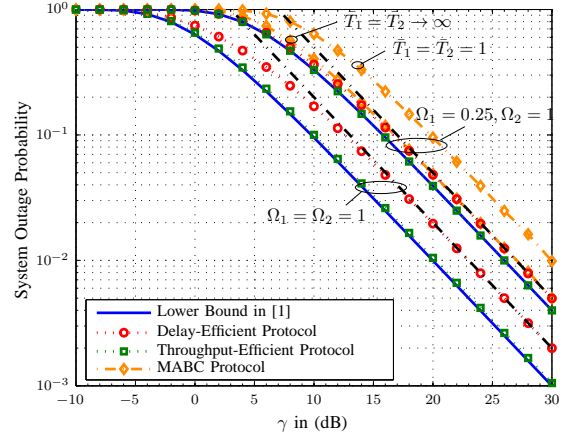}}
\vspace{-0.5cm}
\caption{System outage probability vs. transmit SNR, $\gamma$, in (dB)  for $\ell_1^{\max}=\ell_2^{\max}=10$, $R_0=1$, and the minimum possible average delays. }
\label{FigOutage}
\vspace{-0.5cm}
\end{figure}

In Fig. \ref{FigOutage}, the system outage probability, $F_{\mathrm{sys}}^{\mathrm{out}}$, is plotted versus the transmit SNR, $\gamma$, in (dB). We observe that the system outage probabilities of all considered protocols have diversity order one. Similar to the comparison of the sum throughputs in Fig. \ref{FigRate}, the system outage probability of the proposed protocol in the throughput-efficient mode is lower than that in the delay-efficient mode. Moreover, in the high SNR regime, the SNR gap between the lower bound on the system outage probability given in \cite{ICCIEEE} and the system outage probability of the proposed protocol in the delay-efficient mode is $3$ dB for the symmetric channel and  less than $3$ dB for the asymmetric channel as predicted by (\ref{SNRgap}). Furthermore, the SNR gap between the lower bound and the system outage probability of the proposed protocol in the throughput-efficient mode tends to zero in the high SNR regime. Fig. \ref{FigOutage} shows that, for high SNRs, the proposed protocol in the delay-efficient mode with average delays $(\bar{T}_1,\bar{T}_2)\to(1,1)$ achieves a considerable SNR gain compared to the conventional MABC protocol with average delays $(\bar{T}_1,\bar{T}_2)=(1,1)$. 

\begin{figure}
\centering
\resizebox{0.9\linewidth}{!}{
\psfragfig{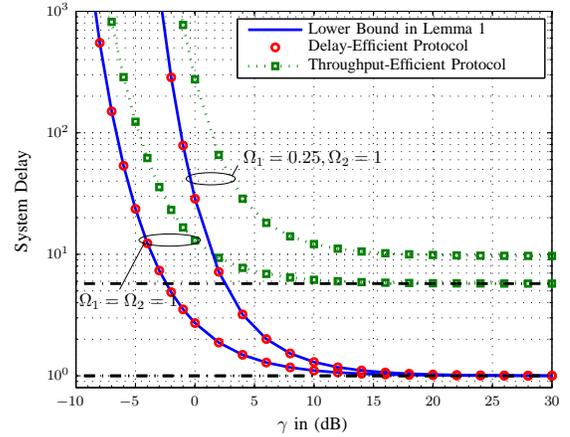}}
\vspace{-0.5cm}
\caption{Minimum possible average system delay vs. transmit SNR, $\gamma$, in (dB)  for $\ell_1^{\max}=\ell_2^{\max}=10$, $R_0=1$.}
\label{FigDelay}
\vspace{-0.5cm}
\end{figure}

In Fig. \ref{FigDelay}, the minimum possible average system delay, $\bar{T}_{\mathrm{sys}}=\frac{\bar{T}_1+\bar{T}_2}{2}$, is depicted versus the transmit SNR, $\gamma$, in (dB). Note that the minimum possible average delay that the proposed protocol in the delay-efficient mode can support is indeed the minimum possible average delay that any adaptive mode selection protocol with causal CSI can achieve, cf. Proposition \ref{PropMinDelayProt1}. Moreover,  the minimum possible average delay of the proposed protocol in the throughput-efficient mode is strictly higher than that of the proposed protocol in the delay-efficient mode.

\section{Conclusion}

In this paper, we proposed a heuristic but efficient delay-constrained protocol with adaptive mode selection for bidirectional relay networks.  The proposed protocol selects a transmission mode in each time slot not only based on the instantaneous qualities of the involved links but also based on the states of the queues at the buffers, i.e., the number of packets in the queues.  Our performance analysis and numerical results revealed that, in the high SNR regime, even for the minimum possible average delay, i.e., one time slot for each information flow, the SNR gap  between the system outage probabilities of the proposed protocol and the delay-unconstrained protocol from \cite{ICCIEEE} is at most $3$ dB. Furthermore, the SNR gap compared to the delay-unconstrained protocol can vanish at the cost of an increased average delay.


\iftoggle{paper}{
  
}{

\appendices

\section{} \label{AppTransProt1} 

\begin{figure*}[!t]
\normalsize
\centering
  \pstool[width=0.9\linewidth]{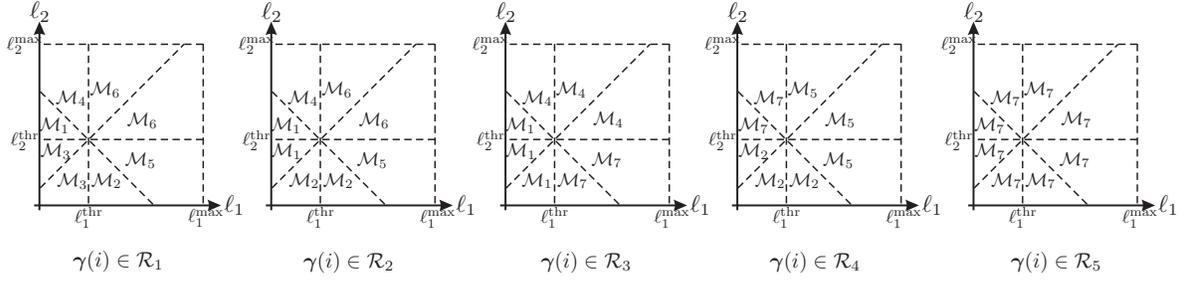}{
\psfrag{M1}[c][c][0.7]{$\mathcal{M}_1$}
\psfrag{M2}[c][c][0.7]{$\mathcal{M}_2$}
\psfrag{M3}[c][c][0.7]{$\mathcal{M}_3$}
\psfrag{M4}[c][c][0.7]{$\mathcal{M}_4$}
\psfrag{M5}[c][c][0.7]{$\mathcal{M}_5$}
\psfrag{M6}[c][c][0.7]{$\mathcal{M}_6$}
\psfrag{M7}[c][c][0.7]{$\mathcal{M}_7$}
\psfrag{l1}[c][c][0.9]{$\ell_1$}
\psfrag{l2}[c][b][0.9]{$\ell_2$}
\psfrag{Lthr1}[c][c][0.7]{$\ell_1^{\mathrm{thr}}$}
\psfrag{Lthr2}[c][c][0.7]{$\ell_2^{\mathrm{thr}}$}
\psfrag{Lmax1}[c][c][0.7]{$\ell_1^{\mathrm{max}}$}
\psfrag{Lmax2}[c][c][0.7]{$\,\,\ell_2^{\mathrm{max}}$}
\psfrag{R1}[c][c][0.8]{$\boldsymbol{\gamma}(i)\in\mathcal{R}_1$}
\psfrag{R2}[c][c][0.8]{$\boldsymbol{\gamma}(i)\in\mathcal{R}_2$}
\psfrag{R3}[c][c][0.8]{$\boldsymbol{\gamma}(i)\in\mathcal{R}_3$}
\psfrag{R4}[c][c][0.8]{$\boldsymbol{\gamma}(i)\in\mathcal{R}_4$}
\psfrag{R5}[c][c][0.8]{$\boldsymbol{\gamma}(i)\in\mathcal{R}_5$}}
\caption{Adaptive mode selection with proposed protocol in the delay-efficient mode based on the states of the queues of the buffers, $\boldsymbol{\ell}(i-1)$, and instantaneous SNRs, $\boldsymbol{\gamma}(i)$.}
\label{FigQRegProof}
\vspace{-0.3cm}
\end{figure*}

For the transition probabilities in (\ref{TransProb_Prot1Eq}a), we refer to Remark \ref{ZeroTransProb}.  In order to obtain the remaining transition probabilities, we use the following partitioning of the possible SNR regions
\begin{IEEEeqnarray}{lll}\label{Partition}  
	m_{s}^{s'} \hspace{-1mm}\Equal \hspace{-1mm} \sum_{m=1}^{5} \hspace{-0.5mm} P_{\mathcal{R}_m} \Pr \left\{\boldsymbol{\ell}(i)=s'|\boldsymbol{\gamma}(i)\in\mathcal{R}_m \wedge \boldsymbol{\ell}(i-1)=s \right\}	 \quad\,\,
\end{IEEEeqnarray}
Note that each of the non-zero transition probabilities provided in Proposition \ref{TransProb_Prot1} corresponds to the probability of selecting one of the transmission modes conditioned on the state of the queues in the previous time slot. For example, transition probability $m_{(\ell_1,\ell_2)}^{(\ell\Add 1,\ell_2)}$ is  the probability of selecting $\mathcal{M}_1$ in the $i$-th time slot conditioned on $\boldsymbol{\ell}(i-1)=(\ell_1,\ell_2)$. In order to simplify the derivation of the transition probabilities, the  transmission modes selected according to the proposed  protocol  based on the states of the queues at the buffers, $\boldsymbol{\ell}(i-1)$, and the instantaneous SNRs, $\boldsymbol{\gamma}(i)$, are illustrated in Fig. \ref{FigQRegProof}. Note that comparing the $\Lambda_k(i)$ for $k=1,\dots,7$ given in (\ref{SelecMet_Prot1}) leads to the boundary conditions for the transmission modes in Fig. \ref{FigQRegProof}, i.e., lines $\ell_1 \Equal 0,\ell_1^{\mathrm{thr}}$, $\ell_2 \Equal 0,\ell_2^{\mathrm{thr}}$,  $\ell_1 \Add \ell_2 \Equal \ell_1^{\mathrm{thr}} \Add \ell_2^{\mathrm{thr}}$, and  $\ell_1 \Minus \ell_2 \Equal \ell_1^{\mathrm{thr}} \Minus \ell_2^{\mathrm{thr}}$. Moreover, the points which are not on the aforementioned lines correspond to the case for which $|\mathcal{U}|=1$ holds. For this case, Fig. \ref{FigQRegProof} illustrates which  transmission modes are selected by the protocol. For the points on the lines, for which $|\mathcal{U}|>1$ holds, the more spectrally efficient mode is selected as stated in the proposed protocol.

For instance for transition probability $m_{(\ell_1,\ell_2)}^{(\ell\Add 1,\ell_2)}$, if  $\ell_2<\ell_2^{\mathrm{thr}}$ and $\ell_1-\ell_2=\ell_1^{\mathrm{thr}}-\ell_2^{\mathrm{thr}}$ hold,  transmission mode $\mathcal{M}_1$ cannot be selected if  $\boldsymbol{\gamma}(i)\in\mathcal{R}_1$ since $\Lambda_1(i)<\Lambda_3(i)$. Furthermore, transmission modes $\mathcal{M}_1$ and $\mathcal{M}_2$ are equiprobable if $\boldsymbol{\gamma}(i)\in\mathcal{R}_2$ since $\Lambda_1(i)=\Lambda_2(i)$ and they have the same spectral efficiency. On the other hand, transmission mode $\mathcal{M}_1$ is selected if $\boldsymbol{\gamma}(i)\in\mathcal{R}_3$ since $\Lambda_1(i)>\Lambda_4(i)$. This leads to $m_{(\ell_1,\ell_2)}^{(\ell\Add 1,\ell_2)}=\frac{P_{\mathcal{R}_2}}{2}+P_{\mathcal{R}_3}$.  All the transition probabilities given in Proposition \ref{TransProb_Prot1} can be obtained in a similar manner. This completes the proof.

\section{} \label{AppProt1MinDel}

The lowest values for the average delays for both information flows are obtained by setting $\ell_1^{\mathrm{thr}}=\ell_2^{\mathrm{thr}}=0$. Therefore, only states $s=(0,0),(1,0),(0,1),(1,1)$ contribute to the analysis of the resulting Markov chain, see Fig. \ref{FigMarkovProof3}. For the relevant states of the Markov chain, using Proposition \ref{TransProb_Prot1}, the following transition probabilities are non-zero
\begin{IEEEeqnarray}{lll} \label{TransProbMinDelayProt1}
		m_{(0,0)}^{(0,0)}= P_{\mathcal{R}_5}, m_{(0,0)}^{(1,0)}= \frac{P_{\mathcal{R}_2}}{2}+P_{\mathcal{R}_3}, m_{(0,0)}^{(0,1)}= \frac{P_{\mathcal{R}_2}}{2}+P_{\mathcal{R}_4}, \nonumber \\
		m_{(0,0)}^{(1,1)}= P_{\mathcal{R}_1}  \IEEEyesnumber\IEEEyessubnumber\\
		m_{(1,0)}^{(0,0)}= P_{\mathcal{R}_1}+P_{\mathcal{R}_2}+P_{\mathcal{R}_4},\,\,  m_{(1,0)}^{(1,0)}= P_{\mathcal{R}_3}+P_{\mathcal{R}_5} \IEEEyessubnumber\\
		m_{(0,1)}^{(0,0)}= P_{\mathcal{R}_1}+P_{\mathcal{R}_2}+P_{\mathcal{R}_3},\,\,  m_{(0,1)}^{(0,1)}= P_{\mathcal{R}_4}+P_{\mathcal{R}_5} \IEEEyessubnumber\\
		m_{(1,1)}^{(0,0)}= P_{\mathcal{R}_1}+P_{\mathcal{R}_2},\,\,  m_{(1,1)}^{(1,0)}= P_{\mathcal{R}_3},\,\,  m_{(1,1)}^{(0,1)}= P_{\mathcal{R}_4},\nonumber \\ m_{(1,1)}^{(1,1)}= P_{\mathcal{R}_5}. \IEEEyessubnumber
\end{IEEEeqnarray}
Using the above transition probabilities in (\ref{SteadyEqu}), we  obtain the state occupancy probabilities as
\begin{IEEEeqnarray}{lll} 
		\Pr\{s=(1,1)\} = \frac{a}{1+a+b+c}  \IEEEyesnumber\IEEEyessubnumber\\
		\Pr\{s=(1,0)\} = \frac{b}{1+a+b+c}  \IEEEyessubnumber\\
		\Pr\{s=(0,1)\} = \frac{c}{1+a+b+c}  \IEEEyessubnumber\\
		\Pr\{s=(0,0)\} = \frac{1}{1+a+b+c},  \IEEEyessubnumber
\end{IEEEeqnarray}
where $a$, $b$, $c$ are given in (\ref{abc}). Substituting the state occupancy and transition probabilities into  (\ref{Throughput}) and (\ref{LittleLaw}), we obtain the average throughputs and the average delays given in (\ref{MinThrProt1}) and (\ref{MinDelayProt1}), respectively. This completes the proof.

\begin{figure}
\normalsize
\centering
 \pstool[width=0.5\linewidth] 
{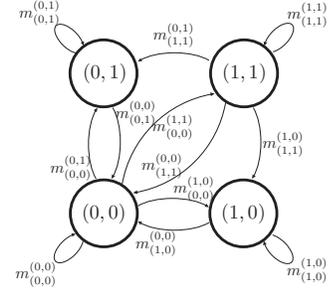}{
\psfrag{A}[c][c][0.75]{$(0,0)$}
\psfrag{B}[c][c][0.75]{$(1,0)$}
\psfrag{C}[c][c][0.75]{$(1,1)$}
\psfrag{D}[c][c][0.75]{$(0,1)$}
\psfrag{m01}[c][c][0.6]{$m_{(0,0)}^{(0,0)}$}
\psfrag{m02}[c][c][0.6]{$m_{(1,0)}^{(0,0)}$}
\psfrag{m03}[c][c][0.6]{$m_{(0,0)}^{(1,0)}$}
\psfrag{m04}[c][c][0.6]{$m_{(1,1)}^{(0,0)}$}
\psfrag{m05}[c][c][0.6]{$m_{(0,0)}^{(1,1)}$}
\psfrag{m06}[c][c][0.6]{$m_{(0,1)}^{(0,0)}$}
\psfrag{m07}[c][c][0.6]{$m_{(0,0)}^{(0,1)}$}
\psfrag{m08}[c][t][0.6]{$m_{(1,0)}^{(1,0)}$}
\psfrag{m09}[c][c][0.6]{$m_{(1,1)}^{(1,0)}$}
\psfrag{m14}[c][c][0.6]{$m_{(1,1)}^{(1,1)}$}
\psfrag{m15}[c][c][0.6]{$m_{(0,1)}^{(0,1)}$}
\psfrag{m16}[c][c][0.6]{$m_{(1,1)}^{(0,1)}$}
}
\caption{ Reduced Markov chain for the number of packets in the queues, $(\ell_1,\ell_2)$, for the proposed protocol in the delay-efficient mode, when $\ell_1^{\mathrm{thr}}=\ell_2^{\mathrm{thr}}=0$. }
\label{FigMarkovProof3}
\vspace{-0.3cm}
\end{figure}
\begin{figure*}[!t]
\normalsize
\centering
 \pstool[width=0.8\linewidth] 
{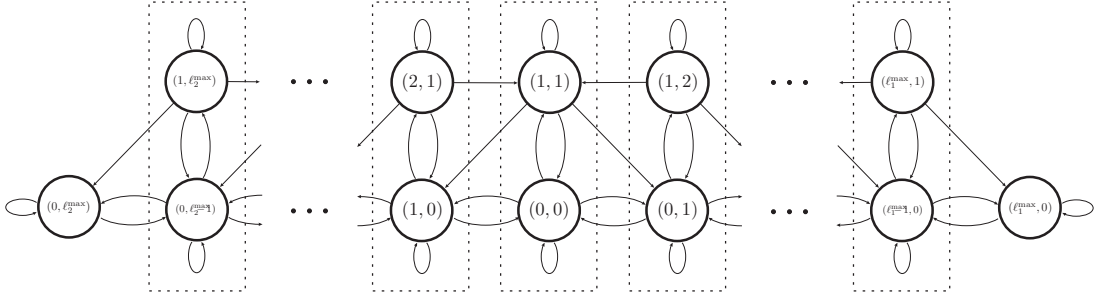}{
\psfrag{A}[c][c][0.7]{$(0,0)$}
\psfrag{B}[c][c][0.7]{$(1,1)$}
\psfrag{C}[c][c][0.7]{$(0,1)$}
\psfrag{D}[c][c][0.7]{$(1,0)$}
\psfrag{E}[c][c][0.7]{$(1,2)$}
\psfrag{F}[c][c][0.7]{$(2,1)$}
\psfrag{G}[c][c][0.4]{$(\ell_1^{\max}\hspace{-5mm}\Minus 1,0)$}
\psfrag{H}[c][c][0.45]{$(\ell_1^{\max},1)$}
\psfrag{I}[c][c][0.45]{$(\ell_1^{\max},0)$}
\psfrag{K}[c][c][0.4]{$(0,\ell_2^{\max}\hspace{-5mm}\Minus1)$}
\psfrag{L}[c][c][0.45]{$(1,\ell_2^{\max})$}
\psfrag{M}[c][c][0.45]{$(0,\ell_2^{\max})$}
}
\caption{ Reduced Markov chain for the number of packets in the queues, $(\ell_1,\ell_2)$, for the proposed protocol in the throughput-efficient when $\ell_1^{\mathrm{thr}}=\ell_2^{\mathrm{thr}}=1$. }
\label{FigMarkovProof2}
\vspace{-0.3cm}
\end{figure*}

\section{} \label{AppProt2Trans}

The non-zero transition probabilities of the proposed protocol in the throughput-efficient mode for $\boldsymbol{\ell}\in\mathcal{L}_2$ and $\boldsymbol{\ell}\in\mathcal{L}_1\cup\mathcal{L}_5\cup\mathcal{L}_6\cup\mathcal{L}_7$ are given in (\ref{TransProbMinDelayProt1}a) and (\ref{TransProbMinDelayProt1}d), respectively. The remaining non-zero transition probabilities  for $\boldsymbol{\ell}\in\mathcal{L}_3$, $\boldsymbol{\ell}\in\mathcal{L}_4$, $\boldsymbol{\ell}\in\mathcal{L}_8$, and $\boldsymbol{\ell}\in\mathcal{L}_9$ are given by
\begin{IEEEeqnarray}{lll}  \label{TransAppRef}
m_{(\ell_1,\ell_2)}^{(\ell_1\Add 1,\ell_2)} \hspace{-0.1mm}\Equal \hspace{-0.5mm} P_{\mathcal{R}_3}, m_{(\ell_1,\ell_2)}^{(\ell_1\Add 1,\ell_2\Add 1)} \hspace{-1mm}\Equal \hspace{-0.5mm} P_{\mathcal{R}_1}, m_{(\ell_1,\ell_2)}^{(\ell_1\Minus 1,\ell_2)} \hspace{-1mm}\Equal \hspace{-0.5mm} P_{\mathcal{R}_2}\Add P_{\mathcal{R}_4},\hspace{-2mm}\nonumber \\ m_{(\ell_1,\ell_2)}^{(\ell_1,\ell_2)} \Equal P_{\mathcal{R}_5}, \IEEEyesnumber\IEEEyessubnumber\\
m_{(\ell_1,\ell_2)}^{(\ell_1\Minus 1,\ell_2)} \Equal P_{\mathcal{R}_1}\Add P_{\mathcal{R}_2}\Add P_{\mathcal{R}_4}, m_{(\ell_1,\ell_2)}^{(\ell_1,\ell_2)} \Equal P_{\mathcal{R}_3}\Add P_{\mathcal{R}_5},  \IEEEyessubnumber \\
m_{(\ell_1,\ell_2)}^{(\ell_1,\ell_2\Minus 1)} \hspace{-0.1mm}\Equal \hspace{-0.5mm} P_{\mathcal{R}_1}\Add P_{\mathcal{R}_2}\Add P_{\mathcal{R}_3}, m_{(\ell_1,\ell_2)}^{(\ell_1,\ell_2)} \Equal P_{\mathcal{R}_4}\Add P_{\mathcal{R}_5},  \IEEEyessubnumber \\
m_{(\ell_1,\ell_2)}^{(\ell_1,\ell_2 \Add 1)} \hspace{-0.1mm}\Equal \hspace{-0.5mm} P_{\mathcal{R}_4},  m_{(\ell_1,\ell_2)}^{(\ell_1\Add 1,\ell_2\Add 1)} \hspace{-1mm}\Equal \hspace{-0.5mm} P_{\mathcal{R}_1},  m_{(\ell_1,\ell_2)}^{(\ell_1,\ell_2\Minus 1)} \hspace{-1mm}\Equal\hspace{-0.5mm} P_{\mathcal{R}_2}\Add P_{\mathcal{R}_3},\hspace{-2mm}\nonumber \\ m_{(\ell_1,\ell_2)}^{(\ell_1,\ell_2)} \Equal P_{\mathcal{R}_5},    \IEEEyessubnumber
\end{IEEEeqnarray}
respectively. With the above transition probabilities, the reduced Markov chain of the number of packets in the queue for the  protocol is illustrated in Fig. \ref{FigMarkovProof2}. Thus, in order to obtain the state occupancy probabilities, we have to solve a system of linear equations with $2(\ell_1^{\max}+\ell_2^{\max})$ unknown variables which leads to the state occupancy probabilities given in (\ref{StateOccProbProt2-TE}). This completes the proof.

\section{} \label{AppProt1High}
Using (\ref{P_Ray}) and (\ref{abc}), for $\gamma\to\infty$, we obtain $P_{\mathcal{R}_1}\to 1$, $P_{\mathcal{R}_m}\to 0,\,\,m=2,3,4,5$, $a\to 1$, and $b,c\to 0$. Hence, we obtain $\bar{R}_{12}\to \frac{R_0}{2}$ and $\bar{R}_{12}\to \frac{R_0}{2}$ from (\ref{MinThrProt1}), and consequently $\bar{R}_{\mathrm{sum}} \to R_0$. In a similar manner, we obtain $(\bar{T}_1^{\mathrm{d}},\bar{T}_2^{\mathrm{d}})\to (1,1)$ from (\ref{MinDelayProt1}). For derivation of the outage probabilities, it is sufficient to substitute the first order approximations of $P_{\mathcal{R}_m},\,\,m=1,\dots,5$ given in (\ref{P_Ray}) and $a=1-\frac{\Omega_1\Omega_2\gamma_{\mathrm{thr}}}{\Omega_1\Omega_2}\cdot\frac{1}{\gamma}+o\left(\frac{1}{\gamma}\right)$, $b=\frac{2\gamma_{\mathrm{thr}}}{\Omega_2}\cdot\frac{1}{\gamma}$, and $c=\frac{2\gamma_{\mathrm{thr}}}{\Omega_1}\cdot\frac{1}{\gamma}$ into (\ref{SysOut}) and simplify the results using the Taylor series $\frac{1}{1+x}=1-x+o(x)$ for $x\to 0$ to obtain (\ref{FoutHighIndPro1}), and consequently the system outage probability in (\ref{HighThrOutPro1}). This completes the proof.

\section{} \label{AppProt2High}
Due to symmetry, we obtain $\bar{R}_{12}=\bar{R}_{12}=\frac{\bar{R}_{\mathrm{sum}}}{2}$ and $F^{\mathrm{out}}_{12}=F^{\mathrm{out}}_{21}=F^{\mathrm{out}}_{\mathrm{sys}}$. In particular, for $\gamma\to\infty$, substituting $P_{\mathcal{R}_1}\to 1$, $P_{\mathcal{R}_m}\to 0,\,\,m=2,3,4,5$ into the state occupancy probabilities given in Proposition \ref{PropMinDelayProt2Modified} leads to 
\begin{IEEEeqnarray}{lll}\label{TransProbProt2ModHigh}
\Pr\{s(\ell_1,\ell_2)\} \hspace{-0.5mm} \Equal \hspace{-0.5mm} \begin{cases}
\frac{1}{2(\ell_1^{\max}+\ell_2^{\max}-1)}, &\hspace{-3mm}\mathrm{if} (\ell_1,\ell_2) \hspace{-0.5mm}\neq \hspace{-0.5mm}(\ell_1^{\max}\hspace{-1mm},0),(0,\ell_2^{\max}) \\
\frac{P_{\mathrm{eq}}}{(\ell_1^{\max}+\ell_2^{\max}-1)}, &\hspace{-3mm}\mathrm{otherwise}
\end{cases} \hspace{-4mm}
\end{IEEEeqnarray}
where $P_{\mathrm{eq}}=P_{\mathcal{R}_3}=P_{\mathcal{R}_4}\to 0$. Hence, we obtain $\bar{R}_{12}\to \frac{R_0}{2}$ and $\bar{R}_{12}\to \frac{R_0}{2}$ from (\ref{Throughput}), and consequently $\bar{R}_{\mathrm{sum}} \to R_0$. For the outage probabilities, we substitute the first order approximations of $P_{\mathcal{R}_m},\,\,m=1,\dots,5$ given in (\ref{P_Ray}) into (\ref{SysOut}) and simplify the results to obtain (\ref{HighThrOutPro2Modified}). The average delays in (\ref{HighDelayPro2Modified}) are  obtained by substituting the transition probabilities (\ref{TransProbProt2ModHigh}) into (\ref{LittleLaw}) and simplifying the results using $\sum_{i=1}^{n}i=\frac{n(n+1)}{2}$. This completes the proof.


}

\bibliographystyle{IEEEtran}
\bibliography{Ref_07_03_2014}

\end{document}